\documentclass[twocolumn]{aastex631}
\usepackage[normalem]{ulem}  %\sout
\usepackage[T1]{fontenc}
\usepackage{ae,aecompl}
\usepackage{graphicx}
\usepackage{amsmath}
\usepackage{amssymb}
\usepackage{supertabular}
\usepackage{hyperref}
\usepackage{xcolor}
\usepackage{academicons}
\usepackage{multirow}
\usepackage{booktabs}
\usepackage{tabularx} 
\usepackage{ragged2e}
\usepackage{threeparttable}
\begin{document}
%--------|---------|---------|---------|---------|---------|---------|---------|
\title{Discovery of Three Glitches in the previously quiet pulsar PSR J1637--4642}

\author{Zhaoyi Wang}
\affiliation{Department of Astronomy, Xiamen University, Xiamen 361005, China; liang@xmu.edu.cn}

\author{Chuwen Zheng}
\affiliation{Department of Astronomy, Xiamen University, Xiamen 361005, China; liang@xmu.edu.cn}

\author{Yuhong Zhuang}
\affiliation{Department of Astronomy, Xiamen University, Xiamen 361005, China; liang@xmu.edu.cn}

\author{Hanyu Zhang}
\affiliation{Department of Astronomy, Xiamen University, Xiamen 361005, China; liang@xmu.edu.cn}

\author{Peng Liu}
\affiliation{Department of Physics and Astronomy, Qinghai University, Xining, 810016, China}
\affiliation{Department of Astronomy, Xiamen University, Xiamen 361005, China; liang@xmu.edu.cn}

\author[0000-0001-6836-9339]{Zhonghao Tu}
\affiliation{Department of Astronomy, Xiamen University, Xiamen 361005, China; liang@xmu.edu.cn}

\author[0000-0001-9849-3656]{Ang Li}
\affiliation{Department of Astronomy, Xiamen University, Xiamen 361005, China; liang@xmu.edu.cn}

%\email{liang@xmu.edu.cn}

\begin{abstract}
We present the discovery and analysis of three rotational glitches in the young pulsar PSR J1637$-$4642.
The timing observations span from 19 February 2009 to 6 October 2024 (MJD 54881--60589) from the Murriyang radio telescope of the Parkes Observatory. The first and strongest glitch occurred around MJD~58352 with a fractional frequency change of $\Delta\nu/\nu \sim 2.7 \times 10^{-6}$, while two additional smaller glitches were detected at MJD~59443 and MJD~60445 with fractional changes of $2.2 \times 10^{-9}$ and $2.8 \times 10^{-8}$, respectively. 
Prior to this, the pulsar had shown no glitch activity since its discovery in the Parkes Multibeam survey.
Only the first glitch exhibits detectable exponential recovery, with a decay timescale of $\sim$100~days and a small recovery fraction $\approx 0.015$, accompanied by a permanent increase in the magnitude of the spin-down rate.
Modeling the post-glitch evolution of $\dot{\nu}$ within the vortex-creep framework using Bayesian inference gives a superfluid moment‑of‑inertia fraction $\approx 0.0187$, consistent with the inner‑crust superfluid.
These results reinforce the standard superfluid glitch paradigm and demonstrate that even ``quiet'' pulsars can still host substantial glitch activity.
\end{abstract}


\section{Introduction}
\label{sec:intro}

Pulsars, rapidly rotating neutron stars emitting beams of electromagnetic radiation, have served as extraordinary laboratories for fundamental physics since their discovery~\citep{1968Natur.217..709H}.
Young pulsars, typically defined as those with characteristic ages $\tau_c < 10^5$ years, are particularly important as they represent the early evolutionary stages of neutron stars and often exhibit complex timing behaviors including glitches and timing noise \citep{2010MNRAS.402.1027H,EspinozaLSKK2011,YuMH2013}.

Glitches are sudden, discrete increases in the rotation frequency of pulsars, typically characterized by fractional amplitudes $\Delta\nu/\nu \sim 10^{-11}$ to $10^{-5}$~\citep{EspinozaLSKK2011,YuMH2013}, superimposed on the steady electromagnetic spin-down, followed by relaxation toward the pre-glitch state.
These phenomena provide a unique window into the interior structure of neutron stars, particularly the superfluid components that are believed to exist in the crust and core~\citep{1972PhRvL..28.1080P,1969Natur.224..872B,1975Natur.256...25A}. 
The superfluid models underpin our understanding of the glitches which involves the transfer of angular momentum between the superfluid neutron vortices and the crustal lattice \citep{1984ApJ...282..533A,1975Natur.256...25A}.

PSR~J1637$-$4642 was discovered in the Parkes Multibeam Pulsar Survey \citep{2003MNRAS.342.1299K}. 
Among the more than 600 pulsars discovered in the survey, PSR J1637$-$4642
is a young isolated pulsar with spin period $P \approx 154$ ms and period derivative $\dot{P} \approx 5.9 \times 10^{-14}$, implying a characteristic age $\tau_c \approx 41$ kyr and a spin-down luminosity $\dot{E} \approx 6.3 \times 10^{35}\,\mathrm{erg \cdot s^{-1}}$~\citep{2010PASA...27...64W}.
While it falls within the position error
contour of the unidentified EGRET gamma-ray source 3EG J1639$-$4702~\citep{2002ASPC..271...31M}, no gamma-ray or X-ray emission has been conclusively detected from this pulsar to date. 
Located in the direction of the supernova remnant Kesteven 41 region, this pulsar exhibits characteristics typical of young, energetic neutron stars, including significant timing noise and evidence for periodic modulation in its pulse arrival times~\citep{2016MNRAS.455.1845K}. 
A useful diagnostic of the long-term rotational evolution of young pulsars is the braking index, $n$. For PSR~J1637$-$4642, a large value of $n = 34(3)$ has been reported \citep{2020MNRAS.494.2012P}; this likely reflects significant timing noise and long-term rotational irregularities.

Despite its youth and high spin-down luminosity---properties typically associated with glitching pulsars \citep{EspinozaLSKK2011,YuMH2013}---PSR~J1637$-$4642 had not exhibited any glitches over a decade since its discovery, although the pulsar has been regularly monitored as part of the Parkes young pulsar timing project \citep{2010PASA...27...64W}.

 In this paper, we report the discovery of three glitches in this pulsar and present detailed timing solutions and physical modeling of the strongest glitch event.
The paper is organized as follows. %Section \ref{sec:observations} describes the Parkes observations and time-of-arrival (ToA) generation. 
Section \ref{sec:observations} describes the observations made with the Murriyang radio telescope, located at Parkes Observatory, and the generation of pulse times of arrival (ToAs).
Section \ref{sec:analysis} presents the timing and glitch models and our fitting procedure. Section \ref{sec:results} reports the measured parameters of the three glitches and models the recovery of the strongest event. Section \ref{sec:discussion} discusses the results in the context of other young glitching pulsars and the implications for superfluid dynamics. Section \ref{sec:conclusion}
 summarizes our conclusions.
 
%%%%%%%%%%%%%%%%%%%%%%%%%%%%%%%%%%%%%%%%%%%%%%%%%
\begin{table*}%[ht]
\small%small%\normalsize%\large
%\begin{minipage}[]{90mm}
\caption{Pre- and post-glitch timing solutions for PSR J1637$-$4642. The timing parameters are obtained by fitting four independent data segments, each bounded by the identified glitch epochs.
\label{tab:int}
} %%%%%%%%%%%%%%%%%%%% table 2 %%%%%%%%%%%%%%%%%%
%\end{minipage}%\vspace{0.2cm}
\begin{flushleft}  %\vspace{-0.2cm}
\hspace*{-\columnsep}  % 去除表格和页面左边的空白，确保左对齐
    \renewcommand{\arraystretch}{1.1}
    \setlength{\tabcolsep}{10pt}    
    \resizebox{1.05 \textwidth}{!}{% 让表格填充页面宽
\begin{tabular}{lcccc}
  \hline   \hline
Parameter                       & Pre-glitch 1  & Post-glitch 1 & Post-glitch 2 & Post-glitch 3  \\ 
\hline
Pulsar name (J2000)             & \multicolumn{4}{c}{PSR J1637$-$4642$^{a}$} \\
Right ascension (J2000) (h:m:s) & \multicolumn{4}{c}{16:37:13.77(6)$^{b}$}\\
Declination (J2000) (d:m:s)     & \multicolumn{4}{c}{$-$46:42:14.2(4)$^{c}$}\\ 
DM (cm$^{-3}$\,pc)              & \multicolumn{4}{c}{419.5(3)$^{b}$}\\ 
\hline
Pulse frequency, $\nu$ (Hz)     & 6.4912942442(4)  & 6.490848760(1) & 6.4905873734(1) & 6.4904642909(1)\\
Pulse frequency derivative, $\dot{\nu}$ ($10^{-12}$\rm\ s$^{-2}$) 
                   & $-$2.495487(4)     & $-$2.50763(8) & $-$2.503100(4)  & $-$2.50375(3)\\
Pulse frequency second derivative, $\ddot{\nu}$ ($10^{-23}$\rm\ s$^{-3}$) 
                  & 3.41(2)            & 10.3(5)  & 2.12(6)  & 24(3) \\
Epoch of frequency determination (MJD) &56601     &58747 &59955 &60525\\
Data span (MJD)   & 54881$-$58321        & 58382$-$59440 & 59454$-$60443 & 60451$-$60590\\
ToA numbers                             &84      &43 &33 &6\\
RMS timing residual ($\rm \mu$s)     &18417     &11822 &937 &36\\
\hline
Time units           &  \multicolumn{4}{c}{TT(TCB)}   \\
Reference time scale &  \multicolumn{4}{c}{TT(TAI)}   \\
Solar System ephemeris model &  \multicolumn{4}{c}{DE440}   \\
  \hline   \hline
\end{tabular} 
}
\end{flushleft}%
\textit{Note}. References: 
$^a$ \citep{2003MNRAS.342.1299K};
$^b$ \citep{2024MNRAS.530.1581K}; 
$^c$ \citep{2019MNRAS.489.3810P}.
\end{table*}
%%%%%%%%%%%%%%%%%%%%%%%%%%%%%%%%%%%%%%%%%%%%%%%%%%%%%%%%%%%%%%%%
%
%%%%%%%%%%%%%%%%%%%% figure 1 %%%%%%%%%%%%%%%%%%
\begin{figure*}
\centering
\includegraphics[width=15 cm]{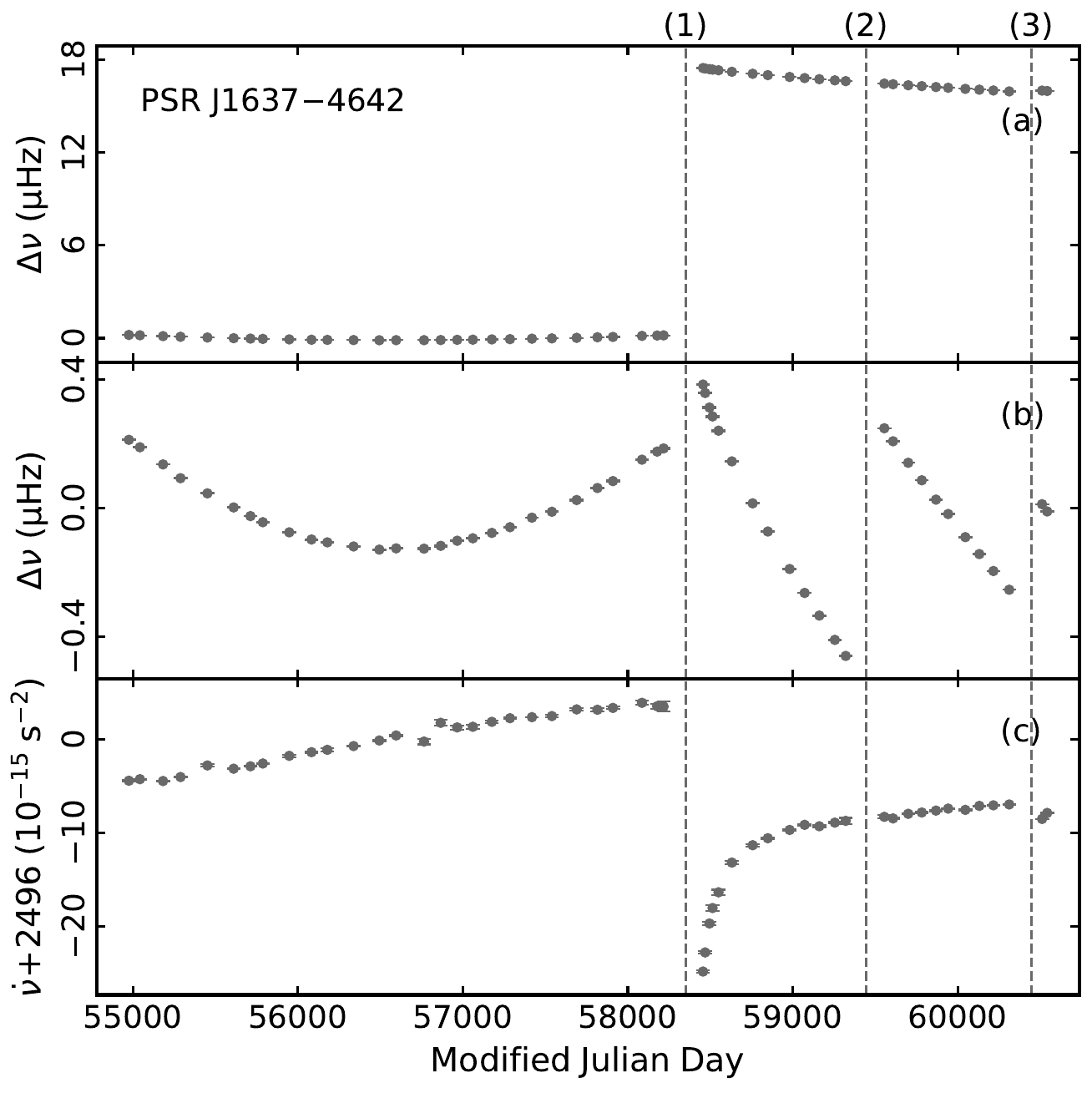}
\caption{
Glitch in PSR J1637$-$4642: 
Panel (a) displays the change of the spin frequency ($\Delta\nu$) relative to the pre-glitch timing model. 
Panel (b) is an enlarged view of the post-glitch $\Delta\nu$ evolution, where each section is centered on its corresponding mean residual value to highlight the recovery details.
Panel (c) shows the temporal evolution trend of the spin frequency derivative ($\dot{\nu}$) before and after the glitch.  
The vertical grey dashed line indicates the glitch epoch, and the numbers in parentheses above denote the corresponding glitch sequence number.
\label{fig:glitch} 
}
\end{figure*}  

\section{Observations and Data Reduction}
\label{sec:observations}

The timing observations of PSR~J1637$-$4642 were conducted using Murriyang, with relevant data archived in the Parkes pulsar data archive\footnote{\href{https://data.csiro.au/domain/atnf}{https://data.csiro.au/domain/atnf}}. 
The timing dataset for PSR~J1637$-$4642 spans from 19 February 2009 to 6 October 2024 (MJD 54881$-$60589). During this period, the pulsar was systematically observed as part of the Parkes young pulsar timing project, with observations occurring approximately every 2$-$4 weeks and integration times of 2$-$15 minutes \citep{JohnstonSDK2021}.

According to the differences in receivers and data acquisition systems, the Parkes timing observations can be broadly divided into three phases:

\textbf{(1) Multi-beam receiver phase (19 February 2009 - 7 February 2020):} Observations were carried out using the multi-beam receiver, with a central frequency of 1369~MHz and a bandwidth of 256~MHz. The data were acquired using the Parkes Digital Filterbank system (PDFB3 and PDFB4) \citep{StaveleyWBD1996, HobbsMDJ2020}. It should be noted that between 13 January and 10 December 2016, the Parkes multi-beam receiver was temporarily replaced by the H-OH receiver, while data acquisition continued with the PDFB4 system. The H-OH receiver operated at a central frequency of 1465~MHz with a bandwidth of 512~MHz, and typical integration times of 10$-$15 minutes \citep{DaiJWK2018}.

\textbf{(2) Dual-receiver phase (7 February 2020 - 23 January 2021):} Starting from 7 February 2020, the Ultra-Wideband Low-frequency (UWL) receiver was introduced, covering a continuous frequency range from 704 to 4032~MHz (with a central frequency of approximately 2368~MHz) \citep{HobbsMDJ2020}. Data acquisition and recording were performed using the Medusa GPU-based signal processing system. During this period, the multi-beam receiver continued to operate simultaneously with the PDFB4 system, enabling parallel recording of narrow-band data.

\textbf{(3) UWL-only phase (after 23 January 2021):} After 23 January 2021, the multi-beam receiver was decommissioned, and all observations were conducted exclusively using the UWL receiver for timing observations.

Based on the raw observational data, we performed offline data reduction using the \texttt{PSRCHIVE} software package \citep{HotanSM2004,StratenDO2012}. 
The detailed procedure is as follows: 
The data were first preprocessed using \texttt{PSRCHIVE} tools, including radio frequency interference (RFI) excision with PAZ and PAZI and incoherent de-dispersion. The median-smoothed difference algorithm is used to identify frequency channels contaminated by RFI, which are then set to zero by PAZ. Residual impulsive interference was identified using PAZI through visual inspection and manually flagged. The incoherent de-dispersion was performed by correcting the frequency-dependent dispersive delays using the measured dispersion measure (DM).
The sub-integrated pulse profiles were then folded and summed over frequency and time to generate the average pulse profile.
For each observing system, the average profile with the highest signal-to-noise ratio and the clearest pulse shape was selected as the reference template, and no smoothing was applied to the pulse profile.
Using the PAT plug-in, pulse ToAs relative to the observatory were obtained by cross-correlating each average profile with its corresponding reference template using the Fourier-domain Markov-chain method \citep{HotanSM2004}.
Owing to the weak radio emission of PSR J1637$-$4642, the pulsar signal was buried in the noise in many observations. As a result, only 166 reliable ToAs were obtained across all observing epochs.

\section{Timing analysis and glitch detection}
\label{sec:analysis}

We next describe the timing procedure used to establish a phase-connected rotational model and to identify glitch signatures in the residuals.

\subsection{Phase Model}

After the initial data reduction, we performed timing analysis on the ToAs to track the rotational evolution of the pulsar. The pulse phase as a function of time is described by a truncated Taylor expansion \citep{EdwardsHM2006}:
\begin{equation}
\label{equ:1}
\phi(t) = \phi_{\rm 0} + \nu(t - t_{\rm 0}) + \frac{\dot{\nu}}{2} (t - t_{\rm 0})^{2} + \frac{\ddot{\nu}}{6}(t - t_{\rm 0})^{3} \ ,
\end{equation}
where $\phi_{\rm 0}$ is the pulse phase at reference epoch $t_{\rm 0}$.
$\nu$, $\dot{\nu}$, $\ddot{\nu}$ are the pulsar spin frequency and its first and second time derivatives, respectively.

We fitted this model to the ToAs using the high-precision pulsar timing software package \texttt{TEMPO2} \citep{EdwardsHM2006,HobbsEM2006}. 
During the fitting process, it is necessary to combine a Solar System ephemeris solution, such as the Jet Propulsion Laboratory  planetary ephemeris DE440 \citep{ParkFWB2021} and Barycentric Coordinate Time (TCB) to convert ToAs to the solar system barycentric reference frame, thereby correcting the influence caused by the Earth motion and local gravity.
Constant phase offsets (“jumps”) between sub-groups of ToAs obtained with different receiver/backend systems are fitted as additional parameters and incorporated into the timing solution. 
To account for underestimated ToA uncertainties, we introduced EFAC and EQUAD parameters using the package \texttt{TEMPO2} following \citet{LiuYGYZ2024}. The ToA effective uncertainty was calculated as $\sigma_{\rm eff}^{2}={\rm EFAC}^{2} \times (\sigma_{\rm ToA}^{2}+{\rm EQUAD}^{2})$, where $\sigma_{\rm ToA}$ represents the initial ToA uncertainty. The EFAC is a free parameter that adjusts unaccounted instrumental errors, and EQUAD accounts for additional time-independent white noise processes~\citep{2014MNRAS.437.3004L}. 

%%%%%%%%%%%%%%%%%%%%%%%%%%%%%%%%%%%%%%%%%%%%%%%%%%%%%%%%%%%%%%%%
\begin{table*}
\large%normalsize
\caption{The timing solutions during the glitch fitting window and glitch parameters obtained for PSR J1637$-$4642.
Timing solutions within the glitch-fitting windows and derived glitch parameters for PSR J1637$-$4642. The recovery fraction $Q$ and decay timescale $\tau_{\rm d}$ are measured only for glitch 1; “$-$” indicates that no significant exponential recovery component is detected.
\label{tab:gl}}  
\begin{flushleft} 
\hspace*{-\dimexpr\columnsep-\arrayrulewidth\relax}
    \renewcommand{\arraystretch}{1.2}
    \setlength{\tabcolsep}{10pt}    
\resizebox{0.99\textwidth}{!}{
\begin{tabular}{lccc}
\hline \hline
Parameter             & Glitch 1    &  Glitch 2  & Glitch 3 \\
\hline
Pulse frequency, $\nu$ (Hz)      
     & 6.490928262(2)   & 6.4907074166(7) & 6.4904911485(4)\\
Pulse frequency derivative, $\dot{\nu}$ ($10^{-12}$\,s$^{-2}$)  
                 & $-$2.4912(2) & $-$2.50405(7) & $-$2.50252(6)\\
Pulse frequency second derivative, $\ddot{\nu}$ ($10^{-23}$\,s$^{-3}$)  
                            & 7.0(7)    & 4.0(4)   & 0.6(4)\\
Epoch of frequency determination (MJD)     
                            & 58300     & 59400    & 60400 \\
Data span (MJD)                            
                  & 57978-59269  & 59066-60024 & 59970-60590\\
ToA numbers                               & 36     & 37  & 21\\
RMS timing residual ($\rm \mu$s)          & 1884   & 778 & 326\\
Glitch epoch (MJD)           & 58352(30)  & 59443(39) & 60445(16)\\
$\Delta\nu$ ($10^{-9}$\,Hz)         & 17543(21) & 14(1)  & 179(1)\\
$\Delta\nu/\nu$ ($10^{-9}$)         & 2703(3)   & 2.2(1) & 27.6(2)\\
$\Delta\dot{\nu}$ ($10^{-14}$\,s$^{-2}$)    
                            & $-$6.4(6)  & $-$0.06(2) & $-$0.13(2)\\
$\Delta\dot{\nu}/\dot{\nu}$ ($10^{-3}$)   
                            & 26(3)      & 0.25(7)    & 0.52(8)\\
$Q$                                 & 0.015(1) & $-$  & $-$\\
$\tau_{\rm d}$ (days)               & 67(8)    & $-$  & $-$\\
\hline  \hline 
\end{tabular}
}
\end{flushleft}%
\vspace{-0.2cm}
\end{table*}
%%%%%%%%%%%%%%%%%%%%%%%%%%%%%%%%%%%%%%%%%%%%%%%%%%%%%%%%%%%%%%%%

\subsection{Glitch Detection}

A glitch introduces an additional contribution to the pulse phase that is not captured by the smooth spin-down model in Eq.~(\ref{equ:1}). In timing residuals, this appears as a sharp change in slope ($\Delta\nu$) and, in some cases, a change in curvature ($\Delta\dot{\nu}$), optionally followed by exponential relaxation. We modelled glitches as step changes in $\nu$ and $\dot{\nu}$ at epoch $t_{\rm g}$, together with an optional decaying frequency term~\citep{EdwardsHM2006}:
%----------------------------------------------------------
\begin{equation}
\begin{split}
\label{equ:2}
\phi_{\rm g} = &\Delta\phi+ \Delta\nu_{\rm p}(t - t_{\rm g}) +   \frac{1} {2} \Delta\dot{\nu}_{\rm p} (t - t_{\rm g})^{2} \\
&+ \sum_{i} {\Delta\nu_{\rm d}^{i}\tau_{\rm d}^{i} [1-e^{-(t - t_{\rm g})/\tau_{\rm d}^{i}}] }\ .
\end{split}
\end{equation} 
%----------------------------------------------------------
Here, $\phi_{\rm g}$ is the phase increment associated with the glitch epoch $t_{\rm g}$. $\Delta\nu_{\rm p}$ and $\Delta\dot{\nu}_{\rm p}$ are the permanent changes in spin frequency and spin-frequency derivative, respectively.
$\Delta\nu_{{\rm d}}^{i}$ and $\tau_{\rm d}^{i}$ are the amplitude and decay timescale of the $i$th transient recovery component.

According to the above equation, the total fractional glitch size is represented as: 
%----------------------------------------------------------
\begin{equation}
\frac{\Delta{\nu}}{{\nu}} = \frac{\Delta{\nu}_{\rm p} + \sum_{\rm  i}\Delta\nu_{\rm d}^{i}} {\nu} \ ,
\end{equation}
%----------------------------------------------------------
For each glitch we fitted $t_{\rm g}$, $\Delta\nu_{\rm p}$, $\Delta\dot{\nu}_{\rm p}$, and any exponential component ($\Delta\nu_{\rm d}^{i}$, $\tau_{\rm d}^{i}$).

%---------------------------------------------------------- 

\subsection{Timing Solutions and Identification of Glitches}

Analysis of the 15.5-year timing dataset from MJD~54881 to MJD~60589 reveals three distinct glitches in PSR~J1637$-$4642. 
The corresponding pre- and post-glitch timing solutions are summarized in Table~\ref{tab:int}. 
We used the \texttt{TEMPO2} package to fit the ToAs based on a timing model including $\nu$, $\dot{\nu}$, and $\ddot{\nu}$. The full dataset was divided into four segments according to the identified glitch epochs, and each segment was analysed independently to obtain coherent timing solutions. The astrometric parameters, dispersion measure, time units, reference time scale, and Solar System ephemeris model were kept fixed during the fitting process, whereas the spin frequency and its first and second derivatives were fitted independently for each data segment.
Since Glitches 1 and 3 introduce phase discontinuities in the timing residuals, we incorporated the parameters of all three glitches into the timing model to achieve phase-connected timing residuals over the full data span %(2009–2024), 
(reported in the Appendix, Figure~\ref{fig:residual}).
Figure~\ref{fig:glitch} shows the long-term timing residuals and spin parameters over the full data span. 
Panel (a) shows the timing residuals relative to the pre-glitch timing solution, with the vertical dashed lines marking the inferred glitch epochs. Panel (b) illustrates the step-like increase in spin frequency at each glitch epoch, while panel (c) shows the accompanying changes in $\dot{\nu}$ and the subsequent post-glitch evolution.
For clarity, zoomed views of the three glitches, showing the timing residuals, spin-frequency evolution, and spin-frequency derivative in the vicinity of each event, are provided in Figures  \ref{fig:glitch1}, \ref{fig:glitch2}, and \ref{fig:glitch3} in the Appendix.

\section{Results and modelling}%%%%%%%%%%%%%%%%%%%%%%%%%%%%%%%%%%%%%%%%%%%
\label{sec:results}

\subsection{Individual Glitch Measurements}

Table~\ref{tab:gl} presents the timing solutions during the glitch fitting window and the glitch parameters obtained for PSR~J1637$-$4642. 
The three glitches show markedly different characteristics:

Glitch 1 was identified as a sudden discontinuity in the rotational phase, frequency, and frequency derivative around MJD~58352. This is by far the strongest of the three glitches, with a fractional frequency change of $\Delta\nu/\nu = 2.703(3) \times 10^{-6}$. Panel~(a) of Figure~\ref{fig:glitch} shows the frequency residuals relative to a pre-glitch timing model, where a sudden jump of approximately $\Delta\nu \approx 17.54$~$\mu$Hz is clearly visible at the glitch epoch. 
This event is followed by a pronounced exponential recovery in $\dot{\nu}$, which we analyze in detail in Secs.~\ref{sec:glitch_1} and \ref{postglitch}.

Glitch 2 occurred at MJD~59443, approximately 3~years after Glitch 1. This glitch is significantly smaller, with a fractional frequency change of $\Delta\nu/\nu = 2.2(1) \times 10^{-9}$, nearly three orders of magnitude smaller than Glitch 1. The frequency jump is $\Delta\nu \approx 14$~nHz. 
No statistically significant exponential recovery was detected for this glitch; the post-glitch timing solution shows a direct transition to the new spin-down rate without measurable relaxation within the sensitivity of the present data.

Glitch 3 occurred at MJD~60445, approximately 2.7~years after Glitch 2 (and about 5 months before the end of our dataset). This glitch is of intermediate size with a fractional frequency change of $\Delta\nu/\nu = 27.6(2) \times 10^{-9}$, about ten times larger than Glitch 2 but still two orders of magnitude smaller than Glitch 1. The frequency jump is $\Delta\nu \approx 179$~nHz. As with Glitch 2, no significant exponential recovery was detected. This non-detection might be limited by sparse post-glitch coverage, with only two data points (derived from six ToAs) available after the event.

The time between the first detection (MJD~54881, 19 Feb 2009) and Glitch 1 (MJD~58352) was approximately 3471~days ($\sim$9.5~years), during which the pulsar was rotationally quiet. 
The recurrence times between the three glitches are irregular:
\begin{itemize}
    \item Between Glitch 1 (MJD~58352) and Glitch 2 (MJD~59443): $\sim$1091~days ($\sim$2.99~years)
    \item Between Glitch 2 (MJD~59443) and Glitch 3 (MJD~60445): $\sim$1002~days ($\sim$2.74~years)
\end{itemize}

\subsection{Apparent Braking Index and Timing Noise}

The long-term spin-down behaviour of PSR~J1637$-$4642 is complex and is relevant for interpreting its glitch activity.
The second frequency derivative $\ddot{\nu}$, obtained from the pre-glitch timing solution, allows us to estimate the apparent braking index $n$ defined as $\dot{\nu} \propto -{\nu}^n$, according to the power-law relation.
By performing a linear fit to the evolution of the spin-down rate before the glitch, we obtain a braking index of $n = 33(1)$, larger than the intrinsic magnetospheric braking of $n = 3$.
A similarly large value, $n=34(3)$, was reported by \citet{2020MNRAS.494.2012P} using approximately 10 years of Murriyang data (MJD~54220--58012; \citealt{2019MNRAS.489.3810P}), which partially overlaps with the data span analysed in this work. The agreement suggests that PSR~J1637$-$4642 exhibits a persistent long-term trend in $\dot{\nu}$ over the available timing baseline.
However, PSR J1637$-$4642 exhibits strong timing noise, which can bias or mimic long-term trends in $\dot{\nu}$~\citep{2023MNRAS.522.4880V,2003A&A...406..667C}. 
A more robust determination of the secular braking index will require dedicated noise modeling beyond the scope of this work.

\subsection{Modeling the Strongest Glitch Event} \label{sec:glitch_1}

Neutron stars are stratified objects with distinct internal regions, each characterized by different physical conditions and compositions~\citep{1983bhwd.book.....S}. The outermost layer consists of a solid crust, approximately 1 km thick, composed of a lattice of atomic nuclei immersed in a degenerate electron gas. Below the crust lies the inner crust, where neutron drip occurs---neutrons begin to escape from nuclei and form a free neutron gas. At even greater depths, the inner crust transitions into the outer core, a region of homogeneous nuclear matter consisting primarily of neutrons with smaller fractions of protons and electrons.

The critical temperature for neutron superfluidity in the inner crust is estimated to be $T_c \sim 10^{9}-10^{10}$ K \citep{2003RvMP...75..607D}. 
Typical neutron star surface temperatures are $T_{\rm s} \sim 10^5-10^6$ K~\citep{2020MNRAS.496.5052P}, and interior temperatures in pulsars older than $\sim100$ yr are several orders of magnitude below $T_{\rm c}$, implying that the inner-crust neutron component should be fully superfluid in PSR J1637$-$4642.

A rotating superfluid cannot spin as a rigid body; instead, it establishes rotation through the formation of quantized vortices.
The standard model for pulsar glitches centers on the interaction between superfluid vortices and the nuclear lattice in the inner crust, mediated by vortex pinning and unpinning processes \citep{1984ApJ...282..533A,1975Natur.256...25A}.
The sudden release of vortices allows the superfluid to transfer angular momentum to the crust, producing the observed glitch, while the post-glitch evolution of the spin-down rate is governed by the creep of superfluid vortices through the inner crust~\citep{1984ApJ...276..325A,1989ApJ...346..823A}.

\begin{figure}
    \centering
    \includegraphics[width=0.48\textwidth]{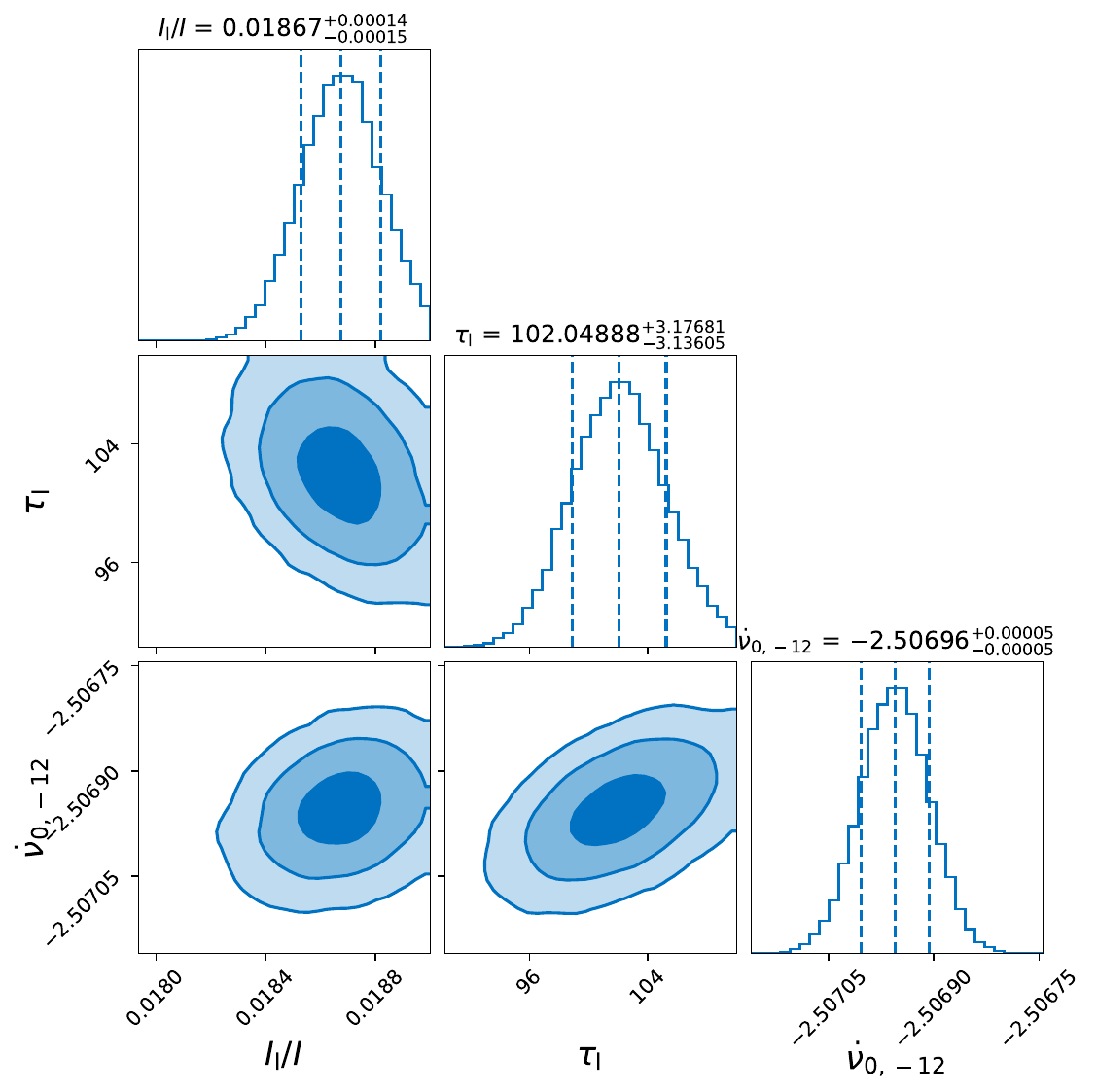}
    \caption{Corner plot showing the posterior distributions and correlations of the vortex creep model parameters. Top panels show 1D histograms with 16th, 50th, and 84th percentiles marked; lower panels show 2D confidence contours at 1$\sigma$ and 2$\sigma$ levels.}
    \label{fig:corner}
\end{figure}

Glitch 1 in PSR J1637$-$4642 is the strongest of the three detected events and the only one exhibiting a clear exponential recovery. Its large amplitude, measurable decay, and significant change in spin-down rate make it particularly suitable for detailed physical modeling. 
The post-glitch evolution of the crustal spin-down rate is given by \citep[e.g.,][]{2025MNRAS.537.1720L}:
\begin{equation}
    \dot{\nu}_{\mathrm{c}}(t) = -\frac{I_{\rm{l}}}{I}\frac{\delta\omega e^{-t/\tau}}{\tau} + \dot{\nu}_0 + \ddot{\nu}t\ ,
    \label{eq:fitting}
\end{equation}
where $\tau$ and $I_{\rm{l}}/I$ are the relaxation time and fractional moment of inertia of the (linear) response region, respectively. 
$\delta\omega$ represents the initial velocity lag between the superfluid and the crust, which in the linear‑response regime is approximately equal to $\Delta\nu$.
$\dot{\nu}_0+\ddot{\nu}t$ is introduced to account for the linear evolution that includes a possible permanent shift $\Delta\dot{\nu}_{\text{per}}$.
The pre-glitch second derivative $\ddot{\nu}$ was obtained by linear fitting of the rotation evolution prior to the glitch. Given that $\ddot{\nu}$ is assumed to be equal before and after the glitch, 
$\Delta\dot{\nu}_{\text{per}}$ can be determined by the difference in the values of the pre‑ and post‑glitch linear evolution at the glitch epoch.

\begin{table*}
\small
\centering
\caption{Comparison of young glitching pulsars. Listed are basic spin parameters and representative glitch properties for six well-studied young pulsars, including PSR J1637$-$4642. From top to bottom, the rows give the spin period $P$, characteristic age $\tau_c = P/(2\dot{P})$, spin-down luminosity $\dot{E}$, surface dipole magnetic field strength $B_{\rm s}$, mean glitch rate (events per year), the typical range of observed fractional glitch sizes $\Delta\nu/\nu$, and representative ranges of recovery fraction $Q$ where available. Where ranges are shown, they are the event ranges reported in the literature.}
\label{tab:comparison}
\renewcommand{\arraystretch}{1.15}
\setlength{\tabcolsep}{2pt}
\begin{tabular}{lcccccc}
\hline\hline
PSR & J0537$-$6910 & Crab  & Vela & \textbf{J1637$-$4642} & J0729$-$1448 & J2337$+$6151 \\
\hline
P (ms) & 16.00$^{a}$ & 33.39$^{b}$ & 89.33$^{c}$ & \textbf{154.07}$^{d}$ & 251.71$^{e}$ & 495.37$^{f}$ \\

$\tau_c$ (kyr) & 4.93 & 1.26 & 11.3 & \textbf{41.1} & 35.2 & 40.6 \\
$\dot{E}$ (10$^{35}$ erg s$^{-1}$) & 4900 & 4500 & 69 & 6.3 & 2.8 & 0.6 \\
%0.63
 $B_{\rm s}$ (10$^{12}$ G) & 0.93 & 3.79 & 3.38 & \textbf{3.06} & 5.41 & 9.91 \\

Rate (yr$^{-1}$) & $\sim$2.41 & $\sim$0.56 & $\sim$0.45 & \textbf{$\sim$0.15} & $\sim$0.23 & $\sim$0.02 \\

$\Delta\nu/\nu (10^{-9})$ & 1.0(3)--687(3)$^{g}$ & 0.8(3)$^{h}$--516.4(1)$^{i}$ & 0.20(2)$^{j}$--3152(2)$^{k}$ & \textbf{2.2(1)--2703(3)} & 3.8(4)$^{m}$--6651.6(8)$^{k}$ &20579.4(12)$^{o}$ \\

 $Q$  & $0.0071(3)^{g}$ &0.6(1)--1.00(4)$^{h}$ & 0.000435(5)$^{l}$--0.1684(4)$^{k}$  & \textbf{0.015(1)} & 0.006(2)$^{i}$ & 0.00751(5)$^{o}$ \\
\hline\hline
\end{tabular}
%\vspace{0.3cm}
\textit{Note}. References: 
$^a$ \citep{2004ApJ...603..682M};
$^b$ \citep{2015MNRAS.446..857L}; 
$^c$ \citep{2002ApJ...564L..85D};
$^d$ \citep{2024MNRAS.530.1581K};
$^e$ \citep{2023ApJ...958..191S};
$^f$ \citep{YuanWM2010};
$^g$ \citep{2018MNRAS.473.1644A};
$^h$ \citep{EspinozaLSKK2011};
$^i$ \citep{BasuSAK2022};
$^j$ \citep{ZubietaGA2024};
$^k$ \citep{YuMH2013};
$^l$ \citep{1988ApJ...330..847C};
$^m$ \citep{BasuJKB2020};
$^o$ \citep{YuanMWZ2010}.
\end{table*}

To fit the vortex-creep recovery model to the post-glitch $\dot\nu(t)$ evolution of Glitch~1, we perform a Bayesian analysis using \texttt{Bilby} \citep{Ashton19}. 
The uniform distributions are adopted as the priors for all fitting parameters, i.e., $I_{\text{l}}/I\sim\text{Unif(0.01, 0.07)}$, $\tau[\text{d}]\sim\text{Unif(50, 200)}$, and  $\dot{\nu}_{0,-12}[\text{Hz/s}]\sim\text{Unif}(-2.510, -2.500)$. 
Here, $\dot{\nu}_{0,-12}$ is $\dot{\nu}_0$ in units of $10^{-12}~\mathrm{Hz/s}$. 
A Gaussian likelihood is adopted for our fitting. The posterior distributions are sampled using the \texttt{dynesty} nested sampler \citep{Speagle20}, which provides robust parameter estimation and evidence calculation.

The analysis yields the following best-fit parameters with their 1$\sigma$ uncertainties: a moment of inertia fraction $I_{\mathrm{l}}/I =0.0187^{+0.0002}_{-0.0001}$ and a relaxation timescale $\tau_{\mathrm{l}} = 102^{+3}_{-3}$~days.
Our analysis also verifies the existence of a permanent shift $\Delta\dot{\nu}_{\text{per}}$ in $\dot{\nu}$, $\Delta\dot{\nu}_{\text{per}}\sim-0.016\times10^{-12}~\mathrm{Hz/s}$.
The 1D and 2D marginalized posterior distributions between model parameters are shown in Figure~\ref{fig:corner}, demonstrating well-constrained parameters inferences.
Previously in Section~\ref{sec:analysis}, \texttt{TEMPO2} was used to derive the timing solution and glitch parameters from the ToAs, adopting the phenomenological exponential recovery model (Eq.~\ref{equ:2}). 
In the Appendix, Figure \ref{fig:residual_vdot}, we compare the physical and phenomenological models, showing that the data provide meaningful support for the more physically motivated model.

It is important to distinguish between the two relaxation timescales derived from our analysis. 
In the standard glitch model used in the timing fit (Eq. \ref{equ:2}), the exponential decay timescale $\tau_{\mathrm{d}}$ describes the phenomenological decay of the observed $\nu$; when interpreted through the physical framework of Eq. \ref{eq:fitting}—which isolates the internal superfluid response from the external torque—we find an intrinsic relaxation timescale of $\tau_{\mathrm{l}} \approx 102$ days; while it was $\tau_{\rm d}=67(8)$ days with the phenomenological fitting model (cf. Table \ref{tab:gl}). The inferred fractional moment of inertia, $I_{\mathrm{l}}/I \approx 1.8\%$, provides a direct estimate of the superfluid reservoir coupled to the crust on these timescales. This value is consistent with theoretical expectations for an inner-crust superfluid origin \citep{1999PhRvL..83.3362L, 2012PhRvL.109x1103A, 2013PhRvL.110a1101C}. 

%%%%%%%%%%%%%%%%%%%%%%%%%%%%%%%%%%%%%%%%%%%%%%%%%%%%%%%%%%%%%%%%
\section{Discussion}
\label{sec:discussion}

PSR J1637$-$4642 adds to the growing class of young pulsars that exhibit large glitches after extended intervals of apparent quiescence. In this section we place these events in the broader context of young glitching pulsars and discuss the constraints they provide on the internal properties of the star.

\subsection{Context: Comparison with Other Glitching Young Pulsars}

Pulsar glitches have now been observed in a substantial and growing population of neutron stars, spanning a broad range of ages, magnetic fields, and spin-down powers. Within this wider population, young pulsars with characteristic ages of $\sim 1$--$100$~kyr and surface magnetic field strengths of order $10^{12}$~G are particularly relevant for comparison with PSR~J1637$-$4642, since they commonly exhibit large glitches and measurable post-glitch recovery. In Table~\ref{tab:comparison}, we therefore compare PSR~J1637$-$4642 with five well-studied young glitching pulsars (PSR~J0537$-$6910, the Crab pulsar, the Vela pulsar, PSR~J0729$-$1448, and PSR~J2337+6151), chosen to span a range of glitch rates, recovery behaviours, and basic spin parameters.

The characteristic age of PSR J1637$-$4642 ($\tau_c \approx 41$ kyr) places it in an intermediate regime between the very young, frequent glitchers ($\tau_c \lesssim 10$ kyr), such as the Crab and Vela, and the more stable population of older pulsars ($\tau_c \gtrsim 100$ kyr), such as PSRs J0729$-$1448 and J 2337+6151. While its glitch rate and overall magnitudes are broadly consistent with its age, PSR J1637$-$4642 shows a remarkably high diversity in glitch size. Its events span over three orders of magnitude—from $\Delta\nu/\nu \approx 2.2 \times 10^{-9}$ (Glitch 2) to $\approx 2.7 \times 10^{-6}$ (Glitch 1).
This range is wide, but it is not unique within the comparison sample: Vela and PSR~J0729$-$1448 also show wide observed glitch-size ranges, and the Crab pulsar exhibits a broad distribution extending from small glitches to events of order $\Delta\nu/\nu\sim10^{-7}$. For PSR~J2337$+$6151, however, only a single representative glitch size is listed, so its event-to-event variability cannot be assessed in the same way. The main distinctive feature of PSR~J1637$-$4642 is that it combines a relatively modest glitch rate with glitches covering more than three orders of magnitude in amplitude.

Within this small comparison set (Table~\ref{tab:comparison}), the pulsars exhibiting the largest fractional glitches have relatively low spin-down luminosity $\dot{E}$. 
The most energetic pulsars, the Crab and PSR J0537$-$6910 ($\dot{E} \gtrsim 4500 \times 10^{35}$erg s$^{-1}$), tend to produce small-to-moderate glitches. 
In contrast, PSR~J1637$-$4642, PSR~J0729$-$1448, and PSR~J2337+6151---which produce the largest glitches ($\Delta\nu/\nu \sim 10^{-6}$ to $2 \times 10^{-5}$)---have the lowest spin-down luminosities ($\dot{E} \approx 0.6-6.3 \times 10^{35}$~erg~s$^{-1}$).
While this may be consistent with larger differential rotation accumulating over longer intervals in slower spin-down systems, a larger and less selected sample is required to test any population-level correlation.

\subsection{Post-Glitch Recovery Patterns and Model Implications} \label{postglitch}

The recovery fraction $Q$ serves as a diagnostic tool for post-glitch relaxation, defined as the ratio of the frequency-decay component to the total glitch size:
\begin{equation}
    Q = \frac{\sum_{i} {\Delta\nu_{\rm d}^{i}} }{\sum_{i} {\Delta\nu_{\rm d}^{i}} + \Delta\nu_{\rm p}} \ .
    \label{eq:Q_ratio}
\end{equation}

For PSR~J1637$-$4642, only Glitch 1 shows detectable exponential recovery, with a recovery fraction $Q = 0.015(1)$, indicating only $\sim$1.5\% decay component and $\sim$98.5\% permanent.
This value is significantly lower than the recovery fractions typically observed in the Crab pulsar ($Q \sim 0.6-1.0$) and is more aligned with the low fractions seen in PSR J0537$-$6910 ($Q \approx 0.007$), PSR J0729$-$1448 ($Q \approx 0.006$), and PSR~J2337+6151 ($Q \approx 0.008$). By comparison, the Vela pulsar exhibits a broad distribution of recovery behavior ($Q \approx 4 \times 10^{-4}$ to $0.17$), occasionally showing much higher partial recovery than what we observe here. The existence of a permanent shift in the spin-down rate, $\Delta\dot{\nu}_{\text{p}}$, further suggests that a fraction of the superfluid angular momentum reservoir was permanently decoupled from the crust during the glitch event.

The post-glitch evolution may offer a direct probe into the underlying vortex creep processes. 
The dominance of linear response in PSR~J1637$-$4642 can be understood from the inferred internal temperature of the neutron star. 
The relaxation timescale $\tau_{\rm l}$ is related to the internal temperature $T$ through the vortex creep activation energy and the pinning potential \citep{1984ApJ...276..325A,2014ApJ...789..141L}: $\tau_\mathrm{l}=\frac{k_{\rm B}T}{E_\mathrm{p}}\frac{\omega_\mathrm{cr}R}{4\Omega_cv_0}\exp(\frac{E_\mathrm{p}}{k_{\rm B}T})$ with the typical velocity $v_0\approx10^5$--$10^7$ cm/s of microscopic vortex motion.
For typical inner crust pinning energies $E_{\rm p} \sim 0.1-1$~MeV~\citep[e.g.,][]{2023PhRvC.108c5808K,2025ApJ...984..200T}, typical neutron star radius $R=12$ km, and assumed critical lag $\omega_{\rm{cr}}\sim10^{-3}$ rad/s, the observed $\tau_{\rm l} \approx 102$~days implies an internal temperature $T \sim 10^8$~K. 
At this temperature, the thermal energy $k_{\rm B} T$ is comparable to the energy
scale of pinning potential variations, allowing vortices to creep gradually without the need for non-linear avalanche processes~\citep{1989ApJ...346..823A}.
The temperature is consistent with the characteristic age of PSR J1637$-$4642, assuming standard neutrino cooling models~\citep[e.g.,][]{1985ApJ...288..191A, 2009PhRvL.102n1101G, 2017MNRAS.469.2313G}. 

\section{Conclusion}
\label{sec:conclusion}

PSR~J1637$-$4642 joins other young pulsars that display large glitches after long quiescent intervals.
It serves as a bridge between the highly active Vela-like pulsars and the more evolved, stable population.
The three glitches we detect span more than three orders of magnitude in fractional size, from $\Delta\nu/\nu \approx 2.2 \times 10^{-9}$ to $ \approx 2.7 \times 10^{-6}$, and show diverse recovery behaviors, with only the strongest event, Glitch~1, displaying a measurable exponential relaxation. 

Modelling the post‑glitch recovery using the vortex-creep framework yields a superfluid moment‑of‑inertia fraction $I_{1}/I\approx 0.0187$, consistent with the inner‑crust superfluid, and an intrinsic relaxation time $\tau_l\approx$ 102 days, which implies an internal temperature $T\sim 10^{8}$ K if pinning energies are $E_{\rm p}\sim$ 0.1$-$1 MeV. 
These results are consistent with the standard superfluid glitch paradigm and the interpretation of the recovery as a linear-creep response. Within this model, the glitch is triggered by the catastrophic unpinning of superfluid vortices from the nuclear lattice, leading to a sudden transfer of angular momentum to the crust. The subsequent exponential recovery reflects the gradual re-pinning and thermal creep of vortices as the system relaxes toward a new equilibrium. The long relaxation timescale and the dominance of the observed linear response are naturally explained by this thermal creep process within the inner-crust superfluid.

Our results demonstrate that even ``quiet'' pulsars can harbor significant glitch activity. 
The long quiescent period before Glitch-1 indicates that stresses can build for many years before a vortex‑unpinning event. 
However, the subsequent detection of three glitches with vastly different sizes in the same pulsar challenges simple stress accumulation models and suggests that glitch triggering depends on complex, possibly stochastic processes in the neutron star interior.
With only three glitches, the recurrence pattern is irregular (separations of 9.5 yr, 3.0 yr, and 2.7 yr); the next event could occur on a timescale of years to decades. 
Several caveats should be kept in mind: only one glitch is suitable for detailed recovery modelling, long-term quantities such as the braking index may be affected by timing noise, and the comparison of its glitch properties with those of other young pulsars is limited by the small number of detected events.
Continued timing will therefore be crucial for measuring the true glitch waiting-time distribution and probing the internal superfluid dynamics.

\section*{Acknowledgments}
We are thankful to the XMU neutron star group for helpful discussions. We acknowledge the support of XMU Training Program of Innovation and Enterpreneurship for Undergraduates.
The work is supported by the National Natural Science Foundation of China (grant Nos. 12273028, 12494572). 
Murriyang, CSIRO’s Parkes radio telescope, is part of the Australia Telescope National Facility, which is funded by the Commonwealth of Australia for operation as a National Facility managed by CSIRO. This paper includes archived data obtained through the CSIRO Data Access Portal.

\appendix

\begin{figure}
\centering
\includegraphics[width=1\textwidth]{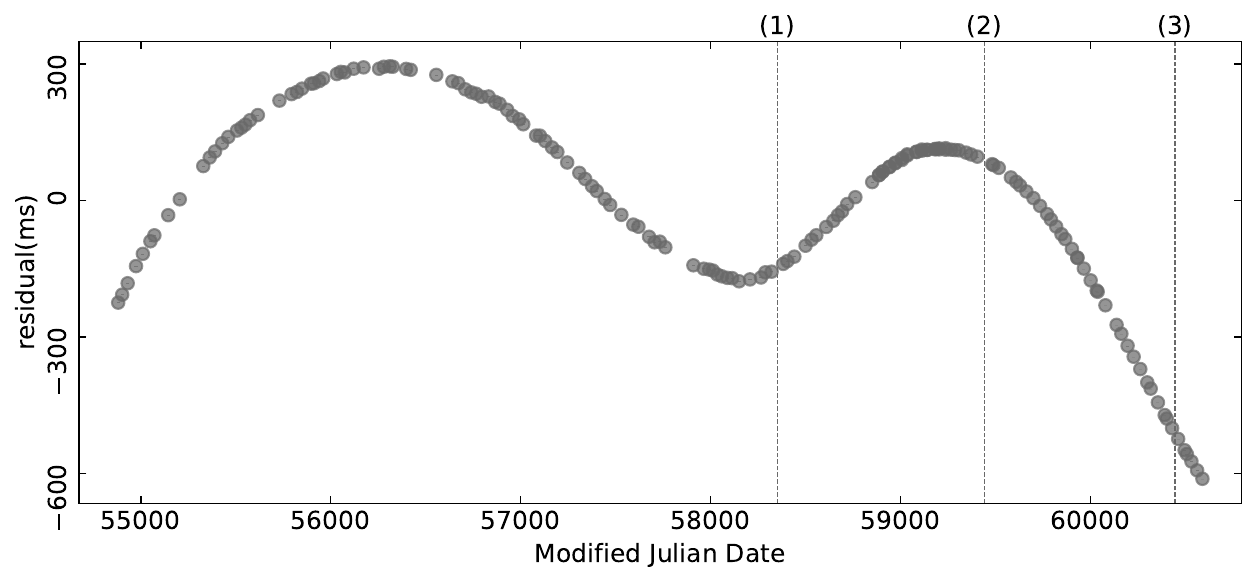}
\caption{
Timing residuals of PSR J1637$-$4642 during MJD 54881--60589 relative to a spin-down model including $\nu$, $\dot{\nu}$, and $\ddot{\nu}$ and the three glitch models, with the glitch model parameters listed in Table~\ref{tab:int}. The vertical dashed lines indicate the glitch epochs, and the numbers in the brackets at the top denote the corresponding glitch sequence numbers.
\label{fig:residual} }
\end{figure}

\begin{figure}
\centering
\includegraphics[width=0.5\textwidth]{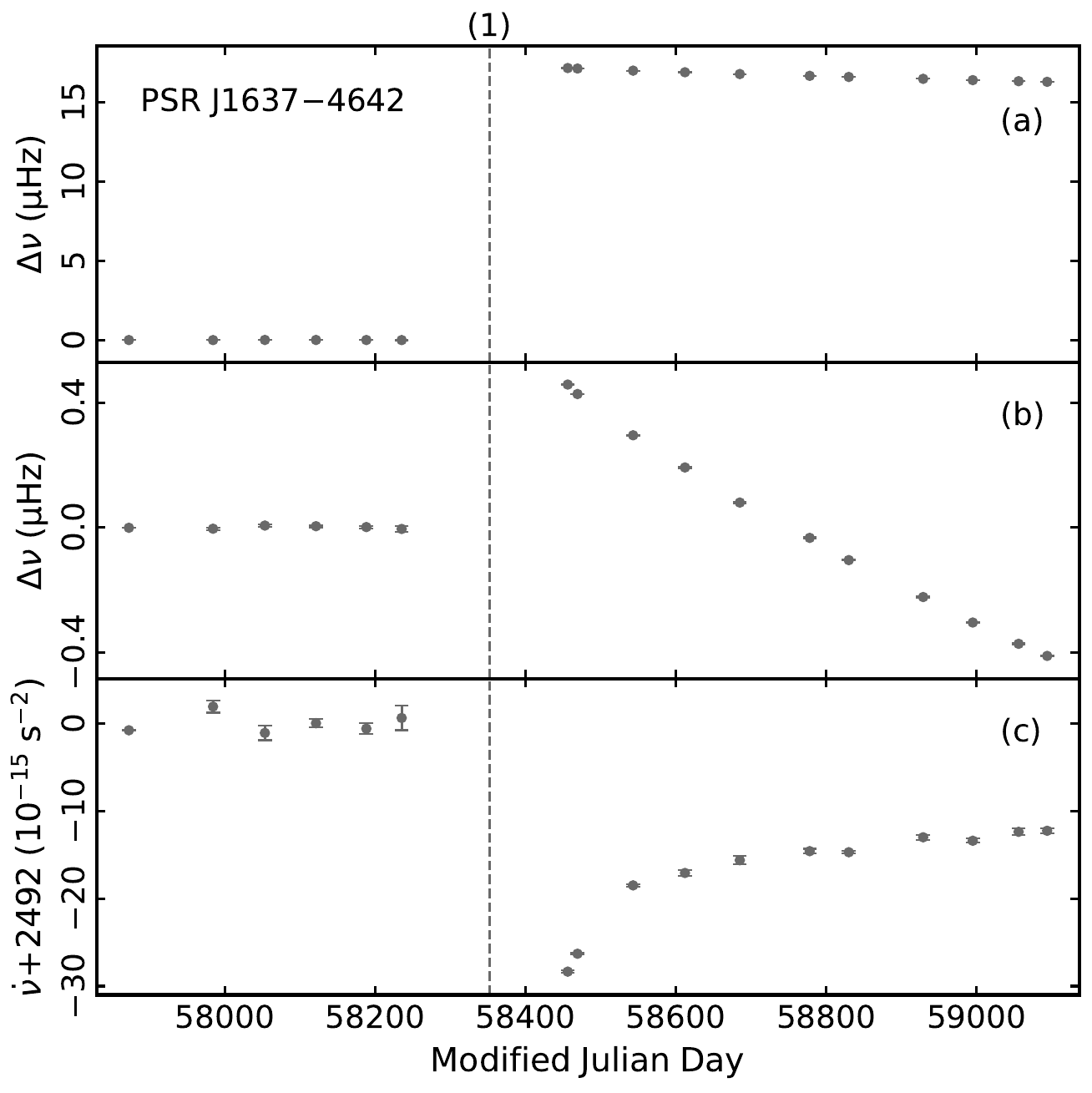}
\caption{
Independent view of glitch~1 in PSR~J1637$-$4642.
Panel (a) shows the timing residuals with respect to the pre-glitch timing model.
Panel (b) shows the evolution of the spin-frequency offset $\Delta\nu$ relative to the pre-glitch solution.
Panel (c) shows the evolution of the spin-frequency derivative $\dot{\nu}$.
The panel definitions and plotting conventions are the same as in Figure~\ref{fig:glitch}, but restricted to the vicinity of glitch~1.
\label{fig:glitch1} }
\end{figure}

\begin{figure}
\centering
\includegraphics[width=0.5\textwidth]{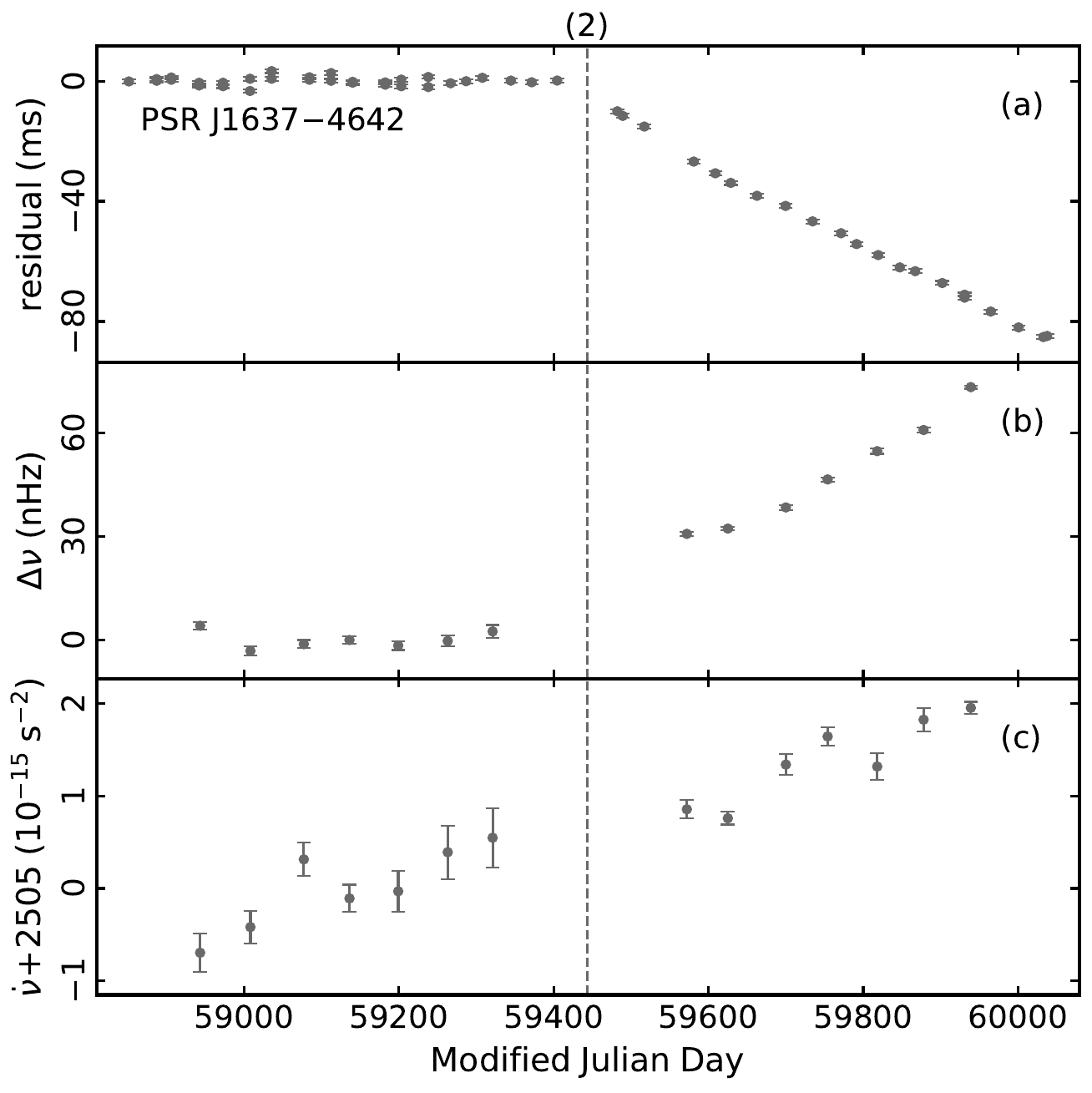}
\caption{
Independent view of glitch~2 in PSR~J1637$-$4642.
Panel (a) presents the timing residuals with respect to the pre-glitch timing model.
Panel (b) shows the evolution of the spin-frequency offset $\Delta\nu$ relative to the pre-glitch solution.
Panel (c) shows the evolution of the spin-frequency derivative $\dot{\nu}$.
The panel definitions and plotting conventions are the same as in Figure~\ref{fig:glitch}, but restricted to the vicinity of glitch~2.
\label{fig:glitch2} 
}
\end{figure}

\begin{figure}
\centering
\includegraphics[width=0.5\textwidth]{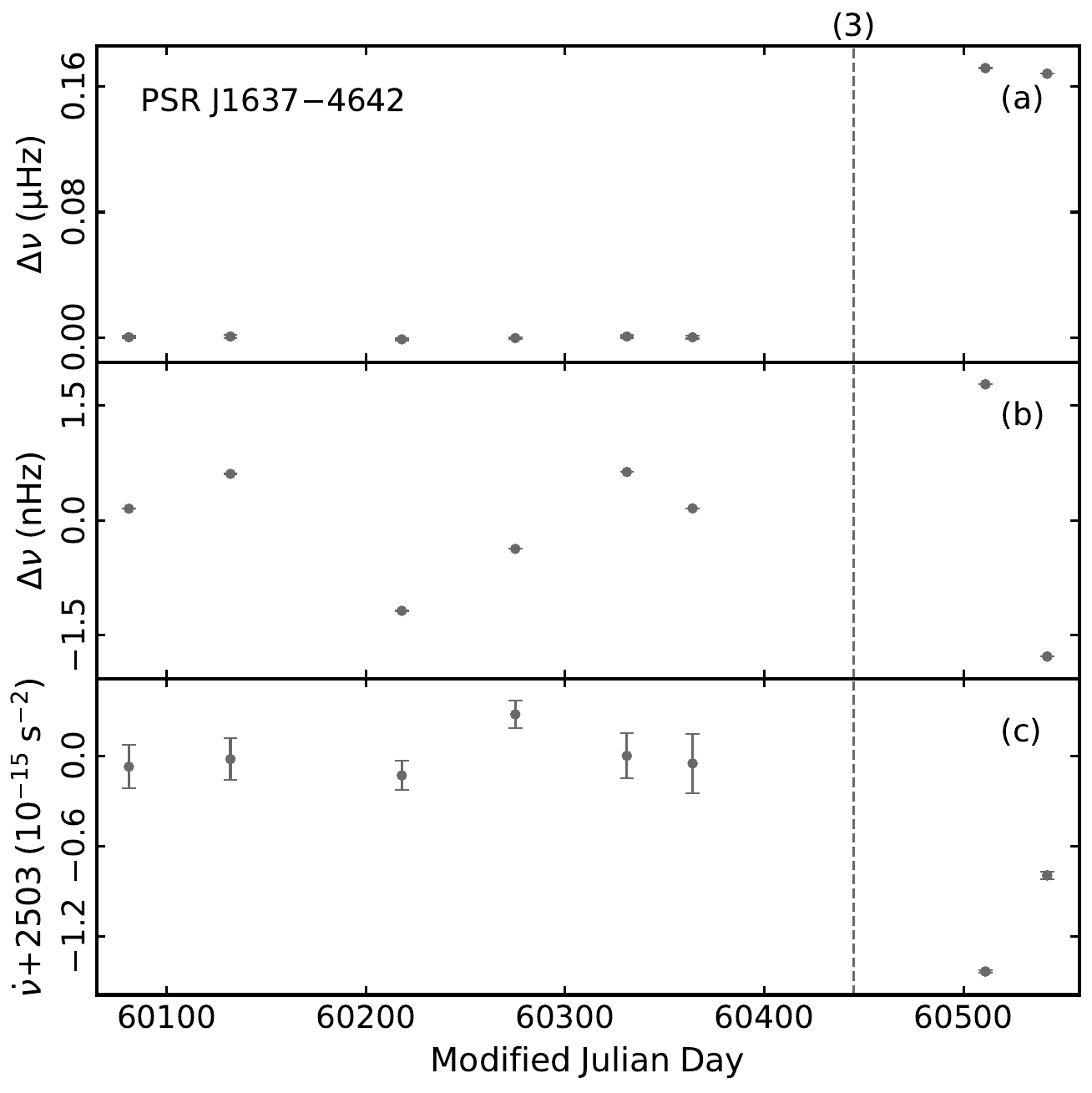}
\caption{
Independent view of glitch~3 in PSR~J1637$-$4642.
Panel (a) shows the timing residuals with respect to the pre-glitch timing model.
Panel (b) shows the evolution of the spin-frequency offset $\Delta\nu$ relative to the pre-glitch solution.
Panel (c) shows the evolution of the spin-frequency derivative $\dot{\nu}$.
The panel definitions and plotting conventions are the same as in Figure~\ref{fig:glitch}, but restricted to the vicinity of glitch~3.
\label{fig:glitch3} }
\end{figure}

\begin{figure}
\centering
\includegraphics[width=0.5\textwidth]{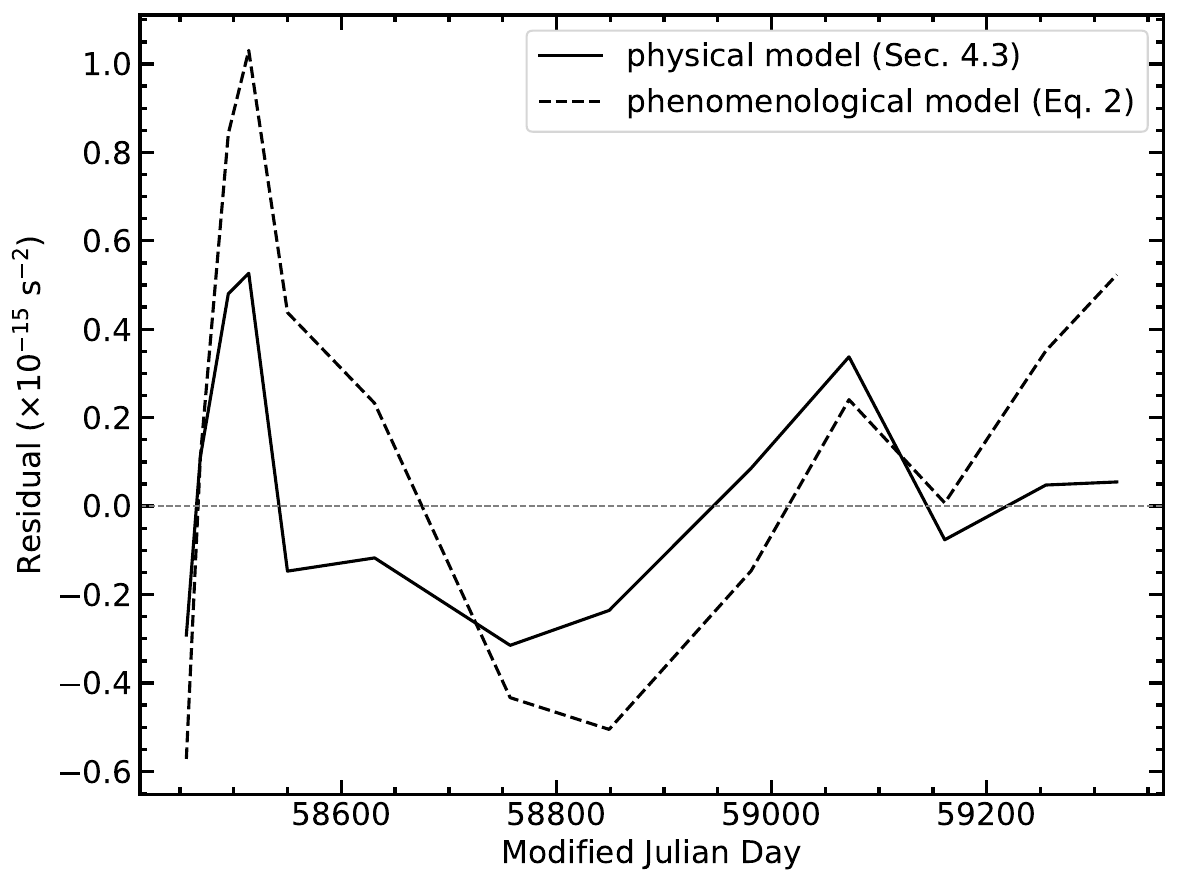}
\caption{
Residuals in $\dot{\nu}$ after fitting the physically motivated model (solid line; Section \ref{sec:glitch_1}) and the phenomenological model (Eq.~\ref{equ:2}; dashed line) to the post-glitch recovery of Glitch~1. 
}\label{fig:residual_vdot}
\end{figure}

Figure~\ref{fig:residual} presents the timing residuals of PSR J1637$-$4642 over the full data span (MJD 54881$-$60589) relative to the spin-down model including $\nu$, $\dot{\nu}$, and $\ddot{\nu}$, after applying the three glitch models.
Separate visualizations of the three glitches are shown in Figures. \ref{fig:glitch1}, \ref{fig:glitch2}, and \ref{fig:glitch3}.
Figure~\ref{fig:residual_vdot} shows the residuals in $\dot{\nu}$ for the post-glitch recovery of Glitch~1 under the phenomenological model, i.e., Eq.~(\ref{equ:2}), and the physically motivated model in Section \ref{sec:glitch_1}. The vortex-creep model provides a better description of the recovery.

\bibliographystyle{aasjournal}
\bibliography{j1637}{}

@ARTICLE{2014MNRAS.437.3004L,
       author = {{Lentati}, L. and {Alexander}, P. and {Hobson}, M.~P. and {Feroz}, F. and {van Haasteren}, R. and {Lee}, K.~J. and {Shannon}, R.~M.},
        title = "{TEMPONEST: a Bayesian approach to pulsar timing analysis}",
      journal = {\mnras},
         year = 2014,
        month = jan,
       volume = {437},
       number = {3},
        pages = {3004-3023},
          doi = {10.1093/mnras/stt2122},
archivePrefix = {arXiv},
       eprint = {1310.2120},
 primaryClass = {astro-ph.IM},
       adsurl = {https://ui.adsabs.harvard.edu/abs/2014MNRAS.437.3004L}
}

@BOOK{1983bhwd.book.....S,
       author = {{Shapiro}, Stuart L. and {Teukolsky}, Saul A.},
        title = "{Black holes, white dwarfs, and neutron stars : the physics of compact objects}",
         year = 1983,
 publisher = {John Wiley \& Sons},
          doi = {10.1002/9783527617661},
       adsurl = {https://ui.adsabs.harvard.edu/abs/1983bhwd.book.....S}
}

@ARTICLE{2020MNRAS.496.5052P,
       author = {{Potekhin}, A.~Y. and {Zyuzin}, D.~A. and {Yakovlev}, D.~G. and {Beznogov}, M.~V. and {Shibanov}, Yu A.},
        title = "{Thermal luminosities of cooling neutron stars}",
      journal = {\mnras},
         year = 2020,
        month = aug,
       volume = {496},
       number = {4},
        pages = {5052-5071},
          doi = {10.1093/mnras/staa1871},
archivePrefix = {arXiv},
       eprint = {2006.15004},
 primaryClass = {astro-ph.HE},
       adsurl = {https://ui.adsabs.harvard.edu/abs/2020MNRAS.496.5052P}
}

@ARTICLE{2014ApJ...789..141L,
       author = {{Link}, Bennett},
        title = "{Thermally Activated Post-glitch Response of the Neutron Star Inner Crust and Core. I. Theory}",
      journal = {\apj},
         year = 2014,
        month = jul,
       volume = {789},
       number = {2},
          eid = {141},
        pages = {141},
          doi = {10.1088/0004-637X/789/2/141},
archivePrefix = {arXiv},
       eprint = {1311.2499},
 primaryClass = {astro-ph.SR},
       adsurl = {https://ui.adsabs.harvard.edu/abs/2014ApJ...789..141L}
}

@ARTICLE{1972PhRvL..28.1080P,
       author = {{Packard}, Richard E.},
        title = "{Pulsar Speedups Related to Metastability of the Superfluid Neutron-Star Core}",
      journal = {\prl},
         year = 1972,
        month = apr,
       volume = {28},
       number = {16},
        pages = {1080-1082},
          doi = {10.1103/PhysRevLett.28.1080},
       adsurl = {https://ui.adsabs.harvard.edu/abs/1972PhRvL..28.1080P}
}

@ARTICLE{1988ApJ...330..847C,
       author = {{Cordes}, J.~M. and {Downs}, G.~S. and {Krause-Polstorff}, J.},
        title = "{JPL Pulsar Timing Observations. V. Macro- and Microjumps in the VELA Pulsar 0833-45}",
      journal = {\apj},
         year = 1988,
        month = jul,
       volume = {330},
        pages = {847},
          doi = {10.1086/166518},
       adsurl = {https://ui.adsabs.harvard.edu/abs/1988ApJ...330..847C}
}

@ARTICLE{2018MNRAS.473.1644A,
       author = {{Antonopoulou}, D. and {Espinoza}, C.~M. and {Kuiper}, L. and {Andersson}, N.},
        title = "{Pulsar spin-down: the glitch-dominated rotation of PSR J0537-6910}",
      journal = {\mnras},
         year = 2018,
        month = jan,
       volume = {473},
       number = {2},
        pages = {1644-1655},
          doi = {10.1093/mnras/stx2429},
archivePrefix = {arXiv},
       eprint = {1708.09459},
 primaryClass = {astro-ph.HE},
       adsurl = {https://ui.adsabs.harvard.edu/abs/2018MNRAS.473.1644A}
}

@ARTICLE{2002ApJ...564L..85D,
       author = {{Dodson}, R.~G. and {McCulloch}, P.~M. and {Lewis}, D.~R.},
        title = "{High Time Resolution Observations of the January 2000 Glitch in the Vela Pulsar}",
      journal = {\apjl},
         year = 2002,
        month = jan,
       volume = {564},
       number = {2},
        pages = {L85-L88},
          doi = {10.1086/339068},
archivePrefix = {arXiv},
       eprint = {astro-ph/0201005},
 primaryClass = {astro-ph},
       adsurl = {https://ui.adsabs.harvard.edu/abs/2002ApJ...564L..85D}
}

@ARTICLE{2023ApJ...958..191S,
       author = {{Smith}, D.~A. and {Abdollahi}, S. and {Ajello}, M. and {Bailes}, M. and {Baldini}, L. and {Ballet}, J. and {Baring}, M.~G. and {Bassa}, C. and {Gonzalez}, J. Becerra and {Bellazzini}, R. and {Berretta}, A. and {Bhattacharyya}, B. and {Bissaldi}, E. and {Bonino}, R. and {Bottacini}, E. and {Bregeon}, J. and {Bruel}, P. and {Burgay}, M. and {Burnett}, T.~H. and {Cameron}, R.~A. and {Camilo}, F. and {Caputo}, R. and {Caraveo}, P.~A. and {Cavazzuti}, E. and {Chiaro}, G. and {Ciprini}, S. and {Clark}, C.~J. and {Cognard}, I. and {Corongiu}, A. and {Orestano}, P. Cristarella and {Crnogorcevic}, M. and {Cuoco}, A. and {Cutini}, S. and {D'Ammando}, F. and {de Angelis}, A. and {DeCesar}, M.~E. and {De Gaetano}, S. and {de Menezes}, R. and {Deneva}, J. and {de Palma}, F. and {Di Lalla}, N. and {Dirirsa}, F. and {Di Venere}, L. and {Dom{\'\i}nguez}, A. and {Dumora}, D. and {Fegan}, S.~J. and {Ferrara}, E.~C. and {Fiori}, A. and {Fleischhack}, H. and {Flynn}, C. and {Franckowiak}, A. and {Freire}, P.~C.~C. and {Fukazawa}, Y. and {Fusco}, P. and {Galanti}, G. and {Gammaldi}, V. and {Gargano}, F. and {Gasparrini}, D. and {Giacchino}, F. and {Giglietto}, N. and {Giordano}, F. and {Giroletti}, M. and {Green}, D. and {Grenier}, I.~A. and {Guillemot}, L. and {Guiriec}, S. and {Gustafsson}, M. and {Harding}, A.~K. and {Hays}, E. and {Hewitt}, J.~W. and {Horan}, D. and {Hou}, X. and {Jankowski}, F. and {Johnson}, R.~P. and {Johnson}, T.~J. and {Johnston}, S. and {Kataoka}, J. and {Keith}, M.~J. and {Kerr}, M. and {Kramer}, M. and {Kuss}, M. and {Latronico}, L. and {Lee}, S.-H. and {Li}, D. and {Li}, J. and {Limyansky}, B. and {Longo}, F. and {Loparco}, F. and {Lorusso}, L. and {Lovellette}, M.~N. and {Lower}, M. and {Lubrano}, P. and {Lyne}, A.~G. and {Maan}, Y. and {Maldera}, S. and {Manchester}, R.~N. and {Manfreda}, A. and {Marelli}, M. and {Mart{\'\i}-Devesa}, G. and {Mazziotta}, M.~N. and {McEnery}, J.~E. and {Mereu}, I. and {Michelson}, P.~F. and {Mickaliger}, M. and {Mitthumsiri}, W. and {Mizuno}, T. and {Moiseev}, A.~A. and {Monzani}, M.~E. and {Morselli}, A. and {Negro}, M. and {Nemmen}, R. and {Nieder}, L. and {Nuss}, E. and {Omodei}, N. and {Orienti}, M. and {Orlando}, E. and {Ormes}, J.~F. and {Palatiello}, M. and {Paneque}, D. and {Panzarini}, G. and {Parthasarathy}, A. and {Persic}, M. and {Pesce-Rollins}, M. and {Pillera}, R. and {Poon}, H. and {Porter}, T.~A. and {Possenti}, A. and {Principe}, G. and {Rain{\`o}}, S. and {Rando}, R. and {Ransom}, S.~M. and {Ray}, P.~S. and {Razzano}, M. and {Razzaque}, S. and {Reimer}, A. and {Reimer}, O. and {Renault-Tinacci}, N. and {Romani}, R.~W. and {S{\'a}nchez-Conde}, M. and {Parkinson}, P.~M. Saz and {Scotton}, L. and {Serini}, D. and {Sgr{\`o}}, C. and {Shannon}, R. and {Sharma}, V. and {Shen}, Z. and {Siskind}, E.~J. and {Spandre}, G. and {Spinelli}, P. and {Stappers}, B.~W. and {Stephens}, T.~E. and {Suson}, D.~J. and {Tabassum}, S. and {Tajima}, H. and {Tak}, D. and {Theureau}, G. and {Thompson}, D.~J. and {Tibolla}, O. and {Torres}, D.~F. and {Valverde}, J. and {Venter}, C. and {Wadiasingh}, Z. and {Wang}, N. and {Wang}, N. and {Wang}, P. and {Weltevrede}, P. and {Wood}, K. and {Yan}, J. and {Zaharijas}, G. and {Zhang}, C. and {Zhu}, W.},
        title = "{The Third Fermi Large Area Telescope Catalog of Gamma-Ray Pulsars}",
      journal = {\apj},
         year = 2023,
        month = dec,
       volume = {958},
       number = {2},
          eid = {191},
        pages = {191},
          doi = {10.3847/1538-4357/acee67},
archivePrefix = {arXiv},
       eprint = {2307.11132},
 primaryClass = {astro-ph.HE},
       adsurl = {https://ui.adsabs.harvard.edu/abs/2023ApJ...958..191S}
}

@ARTICLE{2015MNRAS.446..857L,
       author = {{Lyne}, A.~G. and {Jordan}, C.~A. and {Graham-Smith}, F. and {Espinoza}, C.~M. and {Stappers}, B.~W. and {Weltevrede}, P.},
        title = "{45 years of rotation of the Crab pulsar}",
      journal = {\mnras},
         year = 2015,
        month = jan,
       volume = {446},
       number = {1},
        pages = {857-864},
          doi = {10.1093/mnras/stu2118},
archivePrefix = {arXiv},
       eprint = {1410.0886},
 primaryClass = {astro-ph.HE},
       adsurl = {https://ui.adsabs.harvard.edu/abs/2015MNRAS.446..857L}
}

@ARTICLE{2004ApJ...603..682M,
       author = {{Marshall}, F.~E. and {Gotthelf}, E.~V. and {Middleditch}, J. and {Wang}, Q.~D. and {Zhang}, W.},
        title = "{The Big Glitcher: The Rotation History of PSR J0537-6910}",
      journal = {\apj},
         year = 2004,
        month = mar,
       volume = {603},
       number = {2},
        pages = {682-689},
          doi = {10.1086/381567},
       adsurl = {https://ui.adsabs.harvard.edu/abs/2004ApJ...603..682M}
}

@ARTICLE{2016MNRAS.455.1845K,
       author = {{Kerr}, M. and {Hobbs}, G. and {Johnston}, S. and {Shannon}, R.~M.},
        title = "{Periodic modulation in pulse arrival times from young pulsars: a renewed case for neutron star precession}",
      journal = {\mnras},
         year = 2016,
        month = jan,
       volume = {455},
       number = {2},
        pages = {1845-1854},
          doi = {10.1093/mnras/stv2457},
archivePrefix = {arXiv},
       eprint = {1510.06078},
 primaryClass = {astro-ph.SR},
       adsurl = {https://ui.adsabs.harvard.edu/abs/2016MNRAS.455.1845K}
}

@ARTICLE{2019MNRAS.489.3810P,
       author = {{Parthasarathy}, A. and {Shannon}, R.~M. and {Johnston}, S. and {Lentati}, L. and {Bailes}, M. and {Dai}, S. and {Kerr}, M. and {Manchester}, R.~N. and {Os{\l}owski}, S. and {Sobey}, C. and {van Straten}, W. and {Weltevrede}, P.},
        title = "{Timing of young radio pulsars - I. Timing noise, periodic modulation, and proper motion}",
      journal = {\mnras},
         year = 2019,
        month = nov,
       volume = {489},
       number = {3},
        pages = {3810-3826},
          doi = {10.1093/mnras/stz2383},
archivePrefix = {arXiv},
       eprint = {1908.11709},
 primaryClass = {astro-ph.HE},
       adsurl = {https://ui.adsabs.harvard.edu/abs/2019MNRAS.489.3810P}
}

@ARTICLE{2023MNRAS.522.4880V,
       author = {{Vargas}, Andr{\'e}s F. and {Melatos}, Andrew},
        title = "{Apparent dispersion in pulsar braking index measurements caused by timing noise}",
      journal = {\mnras},
         year = 2023,
        month = jul,
       volume = {522},
       number = {4},
        pages = {4880-4893},
          doi = {10.1093/mnras/stad1301},
archivePrefix = {arXiv},
       eprint = {2305.09079},
 primaryClass = {astro-ph.HE},
       adsurl = {https://ui.adsabs.harvard.edu/abs/2023MNRAS.522.4880V}
}

@ARTICLE{2003MNRAS.342.1299K,
       author = {{Kramer}, M. and {Bell}, J.~F. and {Manchester}, R.~N. and {Lyne}, A.~G. and {Camilo}, F. and {Stairs}, I.~H. and {D'Amico}, N. and {Kaspi}, V.~M. and {Hobbs}, G. and {Morris}, D.~J. and {Crawford}, F. and {Possenti}, A. and {Joshi}, B.~C. and {McLaughlin}, M.~A. and {Lorimer}, D.~R. and {Faulkner}, A.~J.},
        title = "{The Parkes Multibeam Pulsar Survey - III. Young pulsars and the discovery and timing of 200 pulsars}",
      journal = {\mnras},
         year = 2003,
        month = jul,
       volume = {342},
       number = {4},
        pages = {1299-1324},
          doi = {10.1046/j.1365-8711.2003.06637.x},
archivePrefix = {arXiv},
       eprint = {astro-ph/0303473},
 primaryClass = {astro-ph},
       adsurl = {https://ui.adsabs.harvard.edu/abs/2003MNRAS.342.1299K}
}

@ARTICLE{LiuYGYZ2024,
       author = {{Liu}, P. and {Yuan}, J. -P. and {Ge}, M. -Y. and {Ye}, W. -T. and {Zhou}, S. -Q. and {Dang}, S. -J. and {Zhou}, Z. -R. and {G{\"u}gercino{\u{g}}lu}, E. and {Wang}, W. -H. and {Wang}, P. and {Li}, A. and {Li}, D. and {Wang}, N.},
        title = "{Pulse profile variability associated with the glitch of PSR J1048-5832}",
      journal = {\mnras},
         year = 2024,
        month = oct,
       volume = {533},
       number = {4},
        pages = {4274-4286},
          doi = {10.1093/mnras/stae1973},
archivePrefix = {arXiv},
       eprint = {2312.04305},
 primaryClass = {astro-ph.HE},
       adsurl = {https://ui.adsabs.harvard.edu/abs/2024MNRAS.533.4274L}
}

@ARTICLE{ParkFWB2021,
       author = {{Park}, Ryan S. and {Folkner}, William M. and {Williams}, James G. and {Boggs}, Dale H.},
        title = "{The JPL Planetary and Lunar Ephemerides DE440 and DE441}",
      journal = {\aj},
         year = 2021,
        month = mar,
       volume = {161},
       number = {3},
          eid = {105},
        pages = {105},
          doi = {10.3847/1538-3881/abd414},
       adsurl = {https://ui.adsabs.harvard.edu/abs/2021AJ....161..105P}
}

@ARTICLE{HobbsEM2006,
       author = {{Hobbs}, G.~B. and {Edwards}, R.~T. and {Manchester}, R.~N.},
        title = "{TEMPO2, a new pulsar-timing package - I. An overview}",
      journal = {\mnras},
         year = 2006,
        month = jun,
       volume = {369},
       number = {2},
        pages = {655-672},
          doi = {10.1111/j.1365-2966.2006.10302.x},
archivePrefix = {arXiv},
       eprint = {astro-ph/0603381},
 primaryClass = {astro-ph},
       adsurl = {https://ui.adsabs.harvard.edu/abs/2006MNRAS.369..655H}
}

@ARTICLE{EdwardsHM2006,
       author = {{Edwards}, R.~T. and {Hobbs}, G.~B. and {Manchester}, R.~N.},
        title = "{TEMPO2, a new pulsar timing package - II. The timing model and precision estimates}",
      journal = {\mnras},
         year = 2006,
        month = nov,
       volume = {372},
       number = {4},
        pages = {1549-1574},
          doi = {10.1111/j.1365-2966.2006.10870.x},
archivePrefix = {arXiv},
       eprint = {astro-ph/0607664},
 primaryClass = {astro-ph},
       adsurl = {https://ui.adsabs.harvard.edu/abs/2006MNRAS.372.1549E}
}

@ARTICLE{StratenDO2012,
       author = {{van Straten}, Willem and {Demorest}, Paul and {Oslowski}, Stefan},
        title = "{Pulsar Data Analysis with PSRCHIVE}",
      journal = {Astronomical Research and Technology},
         year = 2012,
        month = jul,
       volume = {9},
       number = {3},
        pages = {237-256},
          doi = {10.48550/arXiv.1205.6276},
archivePrefix = {arXiv},
       eprint = {1205.6276},
 primaryClass = {astro-ph.IM},
       adsurl = {https://ui.adsabs.harvard.edu/abs/2012AR&T....9..237V}
}

@ARTICLE{HotanSM2004,
       author = {{Hotan}, A.~W. and {van Straten}, W. and {Manchester}, R.~N.},
        title = "{PSRCHIVE and PSRFITS: An Open Approach to Radio Pulsar Data Storage and Analysis}",
      journal = {\pasa},
         year = 2004,
        month = jan,
       volume = {21},
       number = {3},
        pages = {302-309},
          doi = {10.1071/AS04022},
archivePrefix = {arXiv},
       eprint = {astro-ph/0404549},
 primaryClass = {astro-ph},
       adsurl = {https://ui.adsabs.harvard.edu/abs/2004PASA...21..302H}
}

@ARTICLE{DaiJWK2018,
       author = {{Dai}, S. and {Johnston}, S. and {Weltevrede}, P. and {Kerr}, M. and {Burgay}, M. and {Esposito}, P. and {Israel}, G. and {Possenti}, A. and {Rea}, N. and {Sarkissian}, J.},
        title = "{Peculiar spin frequency and radio profile evolution of PSR J1119-6127 following magnetar-like X-ray bursts}",
      journal = {\mnras},
         year = 2018,
        month = nov,
       volume = {480},
       number = {3},
        pages = {3584-3594},
          doi = {10.1093/mnras/sty2063},
archivePrefix = {arXiv},
       eprint = {1806.05064},
 primaryClass = {astro-ph.HE},
       adsurl = {https://ui.adsabs.harvard.edu/abs/2018MNRAS.480.3584D}
}

@ARTICLE{HobbsMDJ2020,
       author = {{Hobbs}, George and {Manchester}, Richard N. and {Dunning}, Alex and {Jameson}, Andrew and {Roberts}, Paul and {George}, Daniel and {Green}, J.~A. and {Tuthill}, John and {Toomey}, Lawrence and {Kaczmarek}, Jane F. and {Mader}, Stacy and {Marquarding}, Malte and {Ahmed}, Azeem and {Amy}, Shaun W. and {Bailes}, Matthew and {Beresford}, Ron and {Bhat}, N.~D.~R. and {Bock}, Douglas C. -J. and {Bourne}, Michael and {Bowen}, Mark and {Brothers}, Michael and {Cameron}, Andrew D. and {Carretti}, Ettore and {Carter}, Nick and {Castillo}, Santy and {Chekkala}, Raji and {Cheng}, Wan and {Chung}, Yoon and {Craig}, Daniel A. and {Dai}, Shi and {Dawson}, Joanne and {Dempsey}, James and {Doherty}, Paul and {Dong}, Bin and {Edwards}, Philip and {Ergesh}, Tuohutinuer and {Gao}, Xuyang and {Han}, JinLin and {Hayman}, Douglas and {Indermuehle}, Balthasar and {Jeganathan}, Kanapathippillai and {Johnston}, Simon and {Kanoniuk}, Henry and {Kesteven}, Michael and {Kramer}, Michael and {Leach}, Mark and {Mcintyre}, Vince and {Moss}, Vanessa and {Os{\l}owski}, Stefan and {Phillips}, Chris and {Pope}, Nathan and {Preisig}, Brett and {Price}, Daniel and {Reeves}, Ken and {Reilly}, Les and {Reynolds}, John and {Robishaw}, Tim and {Roush}, Peter and {Ruckley}, Tim and {Sadler}, Elaine and {Sarkissian}, John and {Severs}, Sean and {Shannon}, Ryan and {Smart}, Ken and {Smith}, Malcolm and {Smith}, Stephanie and {Sobey}, Charlotte and {Staveley-Smith}, Lister and {Tzioumis}, Anastasios and {van Straten}, Willem and {Wang}, Nina and {Wen}, Linqing and {Whiting}, Matthew},
        title = "{An ultra-wide bandwidth (704 to 4 032 MHz) receiver for the Parkes radio telescope}",
      journal = {\pasa},
         year = 2020,
        month = apr,
       volume = {37},
          eid = {e012},
        pages = {e012},
          doi = {10.1017/pasa.2020.2},
archivePrefix = {arXiv},
       eprint = {1911.00656},
 primaryClass = {astro-ph.IM},
       adsurl = {https://ui.adsabs.harvard.edu/abs/2020PASA...37...12H}
}

@INPROCEEDINGS{2002ASPC..271...31M,
       author = {{Manchester}, R.~N. and {Bell}, J.~F. and {Camilo}, F. and {Kramer}, M. and {Lyne}, A.~G. and {Hobbs}, G.~B. and {Joshi}, B.~C. and {Crawford}, F. and {D'Amico}, N. and {Possenti}, A. and {Kaspi}, V.~M. and {Stairs}, I.~H.},
        title = "{Young Pulsars from the Parkes Multibeam Pulsar Survey and their Associations}",
    booktitle = {Neutron Stars in Supernova Remnants},
         year = 2002,
       editor = {{Slane}, Patrick O. and {Gaensler}, Bryan M.},
       series = {Astronomical Society of the Pacific Conference Series},
       volume = {271},
        month = jan,
        pages = {31},
          doi = {10.48550/arXiv.astro-ph/0112166},
archivePrefix = {arXiv},
       eprint = {astro-ph/0112166},
 primaryClass = {astro-ph},
       adsurl = {https://ui.adsabs.harvard.edu/abs/2002ASPC..271...31M}
}

@ARTICLE{1969Natur.224..872B,
       author = {{Baym}, Gordon and {Pethick}, Christopher and {Pines}, David and {Ruderman}, Malvin},
        title = "{Spin Up in Neutron Stars : The Future of the Vela Pulsar}",
      journal = {\nat},
         year = 1969,
        month = nov,
       volume = {224},
       number = {5222},
        pages = {872-874},
          doi = {10.1038/224872a0},
       adsurl = {https://ui.adsabs.harvard.edu/abs/1969Natur.224..872B}
}

@ARTICLE{1985ApJ...288..191A,
       author = {{Alpar}, M.~A. and {Nandkumar}, R. and {Pines}, D.},
        title = "{Vortex creep and the internal temperature of neutron stars : the Crabpulsar and PSR 0525+21.}",
      journal = {\apj},
         year = 1985,
        month = jan,
       volume = {288},
        pages = {191-195},
          doi = {10.1086/162780},
       adsurl = {https://ui.adsabs.harvard.edu/abs/1985ApJ...288..191A}
}

@ARTICLE{1975Natur.256...25A,
       author = {{Anderson}, P.~W. and {Itoh}, N.},
        title = "{Pulsar glitches and restlessness as a hard superfluidity phenomenon}",
      journal = {\nat},
         year = 1975,
        month = jul,
       volume = {256},
       number = {5512},
        pages = {25-27},
          doi = {10.1038/256025a0},
       adsurl = {https://ui.adsabs.harvard.edu/abs/1975Natur.256...25A}
}

@ARTICLE{1984ApJ...282..533A,
       author = {{Alpar}, M.~A. and {Langer}, S.~A. and {Sauls}, J.~A.},
        title = "{Rapid postglitch spin-up of the superfluid core in pulsars.}",
      journal = {\apj},
         year = 1984,
        month = jul,
       volume = {282},
        pages = {533-541},
          doi = {10.1086/162232},
       adsurl = {https://ui.adsabs.harvard.edu/abs/1984ApJ...282..533A}
}

@article{Speagle20,
    author = {Speagle, J. S.},
    title = "{DYNESTY: a dynamic nested sampling package for estimating Bayesian posteriors and evidences}",
    journal = {Monthly Notices of the Royal Astronomical Society},
    year = 2020,
    volume = 493,
    pages = {3132-3158},
    doi = {10.1093/mnras/staa278}
}

@article{Ashton19,
    author = {Ashton, G. and H{\"u}bner, M. and Lasky, P. D. and others},
    title = "{BILBY: A user-friendly Bayesian inference library for gravitational-wave astronomy}",
    journal = {The Astrophysical Journal Supplement Series},
    year = 2019,
    volume = 241,
    pages = {27},
    doi = {10.3847/1538-4365/ab06fc}
}

@ARTICLE{1999PhRvL..83.3362L,
       author = {{Link}, Bennett and {Epstein}, Richard I. and {Lattimer}, James M.},
        title = "{Pulsar Constraints on Neutron Star Structure and Equation of State}",
      journal = {\prl},
         year = 1999,
        month = oct,
       volume = {83},
       number = {17},
        pages = {3362-3365},
          doi = {10.1103/PhysRevLett.83.3362},
archivePrefix = {arXiv},
       eprint = {astro-ph/9909146},
 primaryClass = {astro-ph},
       adsurl = {https://ui.adsabs.harvard.edu/abs/1999PhRvL..83.3362L}
}

@ARTICLE{2012PhRvL.109x1103A,
       author = {{Andersson}, N. and {Glampedakis}, K. and {Ho}, W.~C.~G. and {Espinoza}, C.~M.},
        title = "{Pulsar Glitches: The Crust is not Enough}",
      journal = {\prl},
         year = 2012,
        month = dec,
       volume = {109},
       number = {24},
          eid = {241103},
        pages = {241103},
          doi = {10.1103/PhysRevLett.109.241103},
archivePrefix = {arXiv},
       eprint = {1207.0633},
 primaryClass = {astro-ph.SR},
       adsurl = {https://ui.adsabs.harvard.edu/abs/2012PhRvL.109x1103A}
}

@ARTICLE{2013PhRvL.110a1101C,
       author = {{Chamel}, N.},
        title = "{Crustal Entrainment and Pulsar Glitches}",
      journal = {\prl},
         year = 2013,
        month = jan,
       volume = {110},
       number = {1},
          eid = {011101},
        pages = {011101},
          doi = {10.1103/PhysRevLett.110.011101},
archivePrefix = {arXiv},
       eprint = {1210.8177},
 primaryClass = {astro-ph.HE},
       adsurl = {https://ui.adsabs.harvard.edu/abs/2013PhRvL.110a1101C}
}

@ARTICLE{2025ApJ...984..200T,
       author = {{Tu}, Zhonghao and {Li}, Ang},
        title = "{Exploring Nuclear Forces with Pulsar Glitch Observations}",
      journal = {\apj},
         year = 2025,
        month = may,
       volume = {984},
       number = {2},
          eid = {200},
        pages = {200},
          doi = {10.3847/1538-4357/adc80e},
archivePrefix = {arXiv},
       eprint = {2412.09219},
 primaryClass = {nucl-th},
       adsurl = {https://ui.adsabs.harvard.edu/abs/2025ApJ...984..200T}
}

@ARTICLE{2023PhRvC.108c5808K,
       author = {{Klausner}, P. and {Barranco}, F. and {Pizzochero}, P.~M. and {Roca-Maza}, X. and {Vigezzi}, E.},
        title = "{Microscopic calculation of the pinning energy of a vortex in the inner crust of a neutron star}",
      journal = {\prc},
         year = 2023,
        month = sep,
       volume = {108},
       number = {3},
          eid = {035808},
        pages = {035808},
          doi = {10.1103/PhysRevC.108.035808},
archivePrefix = {arXiv},
       eprint = {2303.18151},
 primaryClass = {nucl-th},
       adsurl = {https://ui.adsabs.harvard.edu/abs/2023PhRvC.108c5808K}
}

@ARTICLE{2003RvMP...75..607D,
       author = {{Dean}, D.~J. and {Hjorth-Jensen}, M.},
        title = "{Pairing in nuclear systems: from neutron stars to finite nuclei}",
      journal = {Reviews of Modern Physics},
         year = 2003,
        month = apr,
       volume = {75},
       number = {2},
        pages = {607-656},
          doi = {10.1103/RevModPhys.75.607},
archivePrefix = {arXiv},
       eprint = {nucl-th/0210033},
 primaryClass = {nucl-th},
       adsurl = {https://ui.adsabs.harvard.edu/abs/2003RvMP...75..607D}
}

@ARTICLE{1968Natur.217..709H,
       author = {{Hewish}, A. and {Bell}, S.~J. and {Pilkington}, J.~D.~H. and {Scott}, P.~F. and {Collins}, R.~A.},
        title = "{Observation of a Rapidly Pulsating Radio Source}",
      journal = {\nat},
         year = 1968,
        month = feb,
       volume = {217},
       number = {5130},
        pages = {709-713},
          doi = {10.1038/217709a0},
       adsurl = {https://ui.adsabs.harvard.edu/abs/1968Natur.217..709H}
}

@ARTICLE{2010MNRAS.402.1027H,
       author = {{Hobbs}, G. and {Lyne}, A.~G. and {Kramer}, M.},
        title = "{An analysis of the timing irregularities for 366 pulsars}",
      journal = {\mnras},
         year = 2010,
        month = feb,
       volume = {402},
       number = {2},
        pages = {1027-1048},
          doi = {10.1111/j.1365-2966.2009.15938.x},
archivePrefix = {arXiv},
       eprint = {0912.4537},
 primaryClass = {astro-ph.GA},
       adsurl = {https://ui.adsabs.harvard.edu/abs/2010MNRAS.402.1027H}
}

@ARTICLE{2003A&A...406..667C,
       author = {{Chukwude}, A.~E.},
        title = "{On the statistical implication of timing noise for pulsar braking index}",
      journal = {\aap},
         year = 2003,
        month = aug,
       volume = {406},
        pages = {667-671},
          doi = {10.1051/0004-6361:20030789},
       adsurl = {https://ui.adsabs.harvard.edu/abs/2003A&A...406..667C}
}

@ARTICLE{StaveleyWBD1996,
       author = {{Staveley-Smith}, L. and {Wilson}, W.~E. and {Bird}, T.~S. and {Disney}, M.~J. and {Ekers}, R.~D. and {Freeman}, K.~C. and {Haynes}, R.~F. and {Sinclair}, M.~W. and {Vaile}, R.~A. and {Webster}, R.~L. and {Wright}, A.~E.},
        title = "{The Parkes 21 CM multibeam receiver}",
      journal = {\pasa},
         year = 1996,
        month = nov,
       volume = {13},
       number = {3},
        pages = {243-248},
          doi = {10.1017/S1323358000020919},
       adsurl = {https://ui.adsabs.harvard.edu/abs/1996PASA...13..243S}
}

@ARTICLE{JohnstonSDK2021,
       author = {{Johnston}, Simon and {Sobey}, C. and {Dai}, S. and {Keith}, M. and {Kerr}, M. and {Manchester}, R.~N. and {Oswald}, L.~S. and {Parthasarathy}, A. and {Shannon}, R.~M. and {Weltevrede}, P.},
        title = "{Two years of pulsar observations with the ultra-wide-band receiver on the Parkes radio telescope}",
      journal = {\mnras},
         year = 2021,
        month = mar,
       volume = {502},
       number = {1},
        pages = {1253-1262},
          doi = {10.1093/mnras/stab095},
archivePrefix = {arXiv},
       eprint = {2101.07373},
 primaryClass = {astro-ph.HE},
       adsurl = {https://ui.adsabs.harvard.edu/abs/2021MNRAS.502.1253J}
}

@ARTICLE{YuanMWZ2010,
       author = {{Yuan}, J.~P. and {Manchester}, R.~N. and {Wang}, N. and {Zhou}, X. and {Liu}, Z.~Y. and {Gao}, Z.~F.},
        title = "{A Very Large Glitch in PSR B2334+61}",
      journal = {\apjl},
         year = 2010,
        month = aug,
       volume = {719},
       number = {2},
        pages = {L111-L115},
          doi = {10.1088/2041-8205/719/2/L111},
archivePrefix = {arXiv},
       eprint = {1007.1143},
 primaryClass = {astro-ph.HE},
       adsurl = {https://ui.adsabs.harvard.edu/abs/2010ApJ...719L.111Y}
}

@ARTICLE{BasuJKB2020,
       author = {{Basu}, Avishek and {Joshi}, Bhal Chandra and {Krishnakumar}, M.~A. and {Bhattacharya}, Dipankar and {Nandi}, Rana and {Bandhopadhay}, Debades and {Char}, Prasanta and {Manoharan}, P.~K.},
        title = "{Observed glitches in eight young pulsars}",
      journal = {\mnras},
         year = 2020,
        month = jan,
       volume = {491},
       number = {3},
        pages = {3182-3191},
          doi = {10.1093/mnras/stz3230},
archivePrefix = {arXiv},
       eprint = {1911.04934},
 primaryClass = {astro-ph.HE},
       adsurl = {https://ui.adsabs.harvard.edu/abs/2020MNRAS.491.3182B}
}

@ARTICLE{2010PASA...27...64W,
       author = {{Weltevrede}, P. and {Johnston}, S. and {Manchester}, R.~N. and {Bhat}, R. and {Burgay}, M. and {Champion}, D. and {Hobbs}, G.~B. and {K{\i}z{\i}ltan}, B. and {Keith}, M. and {Possenti}, A. and {Reynolds}, J.~E. and {Watters}, K.},
        title = "{Pulsar Timing with the Parkes Radio Telescope for the Fermi Mission}",
      journal = {\pasa},
         year = 2010,
        month = mar,
       volume = {27},
       number = {1},
        pages = {64-75},
          doi = {10.1071/AS09054},
archivePrefix = {arXiv},
       eprint = {0909.5510},
 primaryClass = {astro-ph.GA},
       adsurl = {https://ui.adsabs.harvard.edu/abs/2010PASA...27...64W}
}

@ARTICLE{2024MNRAS.530.1581K,
       author = {{Keith}, M.~J. and {Johnston}, S. and {Karastergiou}, A. and {Weltevrede}, P. and {Lower}, M.~E. and {Basu}, A. and {Posselt}, B. and {Oswald}, L.~S. and {Parthasarathy}, A. and {Cameron}, A.~D. and {Serylak}, M. and {Buchner}, S.},
        title = "{The Thousand-Pulsar-Array programme on MeerKAT ─ XIII. Timing, flux density, rotation measure, and dispersion measure time series of 597 pulsars}",
      journal = {\mnras},
         year = 2024,
        month = may,
       volume = {530},
       number = {2},
        pages = {1581-1591},
          doi = {10.1093/mnras/stae937},
archivePrefix = {arXiv},
       eprint = {2404.02051},
 primaryClass = {astro-ph.HE},
       adsurl = {https://ui.adsabs.harvard.edu/abs/2024MNRAS.530.1581K}
}

@ARTICLE{EspinozaLSKK2011,
       author = {{Espinoza}, C.~M. and {Lyne}, A.~G. and {Stappers}, B.~W. and {Kramer}, M.},
        title = "{A study of 315 glitches in the rotation of 102 pulsars}",
      journal = {\mnras},
         year = 2011,
        month = jun,
       volume = {414},
       number = {2},
        pages = {1679-1704},
          doi = {10.1111/j.1365-2966.2011.18503.x},
archivePrefix = {arXiv},
       eprint = {1102.1743},
 primaryClass = {astro-ph.HE},
       adsurl = {https://ui.adsabs.harvard.edu/abs/2011MNRAS.414.1679E}
}

@ARTICLE{BasuSAK2022,
       author = {{Basu}, A. and {Shaw}, B. and {Antonopoulou}, D. and {Keith}, M.~J. and {Lyne}, A.~G. and {Mickaliger}, M.~B. and {Stappers}, B.~W. and {Weltevrede}, P. and {Jordan}, C.~A.},
        title = "{The Jodrell bank glitch catalogue: 106 new rotational glitches in 70 pulsars}",
      journal = {\mnras},
         year = 2022,
        month = mar,
       volume = {510},
       number = {3},
        pages = {4049-4062},
          doi = {10.1093/mnras/stab3336},
archivePrefix = {arXiv},
       eprint = {2111.06835},
 primaryClass = {astro-ph.HE},
       adsurl = {https://ui.adsabs.harvard.edu/abs/2022MNRAS.510.4049B}
}

@article{YuMH2013,
  author = {{Yu}, M. and {Manchester}, R. N. and {Hobbs}, G. and {Johnston}, S. and {Kaspi}, V. M. and {Keith}, M. and {Lyne}, A. G. and {Qiao}, G. J. and {Ravi}, V. and {Sarkissian}, J. M. and {Shannon}, R. and {Xu}, R. X.},
  title = {Detection of 107 glitches in 36 southern pulsars},
  journal = {mnras},
  year = {2013},
  month = {February},
  volume = {429},
  pages = {688-724},
  doi = {10.1093/mnras/sts366},
  url = {http://adsabs.harvard.edu/abs/2013MNRAS.429..688Y}
}

@ARTICLE{ZubietaGA2024,
       author = {{Zubieta}, E. and {Garc{\'\i}a}, F. and {del Palacio}, S. and {Araujo Furlan}, S.~B. and {Gancio}, G. and {Lousto}, C.~O. and {Combi}, J.~A. and {Espinoza}, C.~M.},
        title = "{Timing irregularities and glitches from the pulsar monitoring campaign at IAR}",
      journal = {\aap},
         year = 2024,
        month = sep,
       volume = {689},
          eid = {A191},
        pages = {A191},
          doi = {10.1051/0004-6361/202450441},
archivePrefix = {arXiv},
       eprint = {2406.17099},
 primaryClass = {astro-ph.HE},
       adsurl = {https://ui.adsabs.harvard.edu/abs/2024A&A...689A.191Z}
}

@ARTICLE{YuanWM2010,
       author = {{Yuan}, J.~P. and {Wang}, N. and {Manchester}, R.~N. and {Liu}, Z.~Y.},
        title = "{29 glitches detected at Urumqi Observatory}",
      journal = {\mnras},
         year = 2010,
        month = may,
       volume = {404},
       number = {1},
        pages = {289-304},
          doi = {10.1111/j.1365-2966.2010.16272.x},
archivePrefix = {arXiv},
       eprint = {1001.1471},
 primaryClass = {astro-ph.GA},
       adsurl = {https://ui.adsabs.harvard.edu/abs/2010MNRAS.404..289Y}
}

@ARTICLE{1984ApJ...276..325A,
       author = {{Alpar}, M.~A. and {Anderson}, P.~W. and {Pines}, D. and {Shaham}, J.},
        title = "{Vortex creep and the internal temperature of neutron stars. I - General theory}",
      journal = {\apj},
         year = 1984,
        month = jan,
       volume = {276},
        pages = {325-334},
          doi = {10.1086/161616},
       adsurl = {https://ui.adsabs.harvard.edu/abs/1984ApJ...276..325A}
}

@ARTICLE{1989ApJ...346..823A,
       author = {{Alpar}, M.~A. and {Cheng}, K.~S. and {Pines}, D.},
        title = "{Vortex Creep and the Internal Temperature of Neutron Stars: Linear and Nonlinear Response to a Glitch}",
      journal = {\apj},
         year = 1989,
        month = nov,
       volume = {346},
        pages = {823},
          doi = {10.1086/168063},
       adsurl = {https://ui.adsabs.harvard.edu/abs/1989ApJ...346..823A}
}

@ARTICLE{2025MNRAS.537.1720L,
       author = {{Liu}, P. and {Yuan}, J.-P. and {Ge}, M.-Y. and {Ye}, W.-T. and {Zhou}, S.-Q. and {Dang}, S.-J. and {Zhou}, Z.-R. and {G{\"u}gercino{\u{g}}lu}, E. and {Tu}, Z.-H. and {Wang}, P. and {Li}, A. and {Li}, D. and {Wang}, N.},
        title = "{A multiband study of pulsar glitches with Fermi-LAT and Parkes}",
      journal = {\mnras},
         year = 2025,
        month = feb,
       volume = {537},
       number = {2},
        pages = {1720-1734},
          doi = {10.1093/mnras/staf101},
archivePrefix = {arXiv},
       eprint = {2408.15022},
 primaryClass = {astro-ph.HE},
       adsurl = {https://ui.adsabs.harvard.edu/abs/2025MNRAS.537.1720L}
}

@ARTICLE{2020MNRAS.494.2012P,
       author = {{Parthasarathy}, A. and {Johnston}, S. and {Shannon}, R.~M. and {Lentati}, L. and {Bailes}, M. and {Dai}, S. and {Kerr}, M. and {Manchester}, R.~N. and {Os{\l}owski}, S. and {Sobey}, C. and {van Straten}, W. and {Weltevrede}, P.},
        title = "{Timing of young radio pulsars - II. Braking indices and their interpretation}",
      journal = {\mnras},
         year = 2020,
        month = may,
       volume = {494},
       number = {2},
        pages = {2012-2026},
          doi = {10.1093/mnras/staa882},
archivePrefix = {arXiv},
       eprint = {2003.13303},
 primaryClass = {astro-ph.HE},
       adsurl = {https://ui.adsabs.harvard.edu/abs/2020MNRAS.494.2012P}
}

@ARTICLE{2009PhRvL.102n1101G,
       author = {{Glampedakis}, Kostas and {Andersson}, Nils},
        title = "{Hydrodynamical Trigger Mechanism for Pulsar Glitches}",
      journal = {\prl},
         year = 2009,
        month = apr,
       volume = {102},
       number = {14},
          eid = {141101},
        pages = {141101},
          doi = {10.1103/PhysRevLett.102.141101},
archivePrefix = {arXiv},
       eprint = {0806.3664},
 primaryClass = {astro-ph},
       adsurl = {https://ui.adsabs.harvard.edu/abs/2009PhRvL.102n1101G}
}

@ARTICLE{2017MNRAS.469.2313G,
       author = {{G{\"u}gercino{\u{g}}lu}, Erbil},
        title = "{Post-glitch exponential relaxation of radio pulsars and magnetars in terms of vortex creep across flux tubes}",
      journal = {\mnras},
         year = 2017,
        month = aug,
       volume = {469},
       number = {2},
        pages = {2313-2322},
          doi = {10.1093/mnras/stx985},
archivePrefix = {arXiv},
       eprint = {1701.05786},
 primaryClass = {astro-ph.HE},
       adsurl = {https://ui.adsabs.harvard.edu/abs/2017MNRAS.469.2313G}
}

\end{document}